\documentclass[pra,showkeys,floatfix,twocolumn]{revtex4-2}
\usepackage{graphicx}
\usepackage[utf8]{inputenc}
\usepackage{amsmath}
\usepackage{amssymb}
\usepackage{color}
\usepackage{tabularx}

\begin{document}

\title{Optical vortex bullets with large angular momenta}
\author{Jos\'{e} R. Salgueiro$^{1}$, Carlos F. S\'{a}nchez$^{1,2}$, Boris A.
Malomed$^{3,4}$, \'{A}ngel Paredes$^{1}$, and Humberto Michinel$^{1}$}
\affiliation{
$^1$Instituto de F\'isica, Computaci\'on e Ciencia Aeroespacial (IFCAE),
Universidade de Vigo.\\ Campus de As Lagoas, E--32004 Ourense, Spain.\\
$^2$Grupo de Investigaci\'on en F\'isica (GIF).
Universidad Polit\'ecnica Salesiana, Cuenca, 010105, Ecuador.\\
$^3$ Department of Physical Electronics, School of Electric and Computer
Engineering, Faculty of Engineering, Tel Aviv University, Tel Aviv, Israel\\
$^4$Instituto de Alta Investigaci\'{o}n, Universidad de Tarapac\'{a}, Casilla 7D, Arica, Chile}

\begin{abstract}
%----------------------------   ABSTRACT  ------------------------------------

We investigate three-dimensional spatiotemporal optical vortex solitons
(light bullets with a toroidal shape) carrying higher-order
orbital angular momentum in Kerr media with anomalous group-velocity
dispersion and a transversely localized Gaussian profile of the linear
refractive index. Stationary vortex bullets with topological charges
(winding numbers) $1\leq \ell \leq 4$ are obtained as numerical solutions of
the three-dimensional nonlinear Schr\"{o}dinger equation, and their
existence, structure and stability are systematically analyzed. In terms of atomic
Bose-Einstein condensates, these solutions represent matter-wave solitons
with embedded vorticity. We examine the dependence of the solution families
on the propagation constant, the depth and width of the guiding profile, and
$\ell $. It is found that the increase of $\ell $ requires one to use a
stronger or wider waveguiding potential to sustain localized states
satisfying the Vakhitov-Kolokolov (VK) stability criterion. VK-complying and
non-complying branches of vortex-soliton families are
separated by a degenerate Townes branch. Direct simulations reveal that,
although the guiding profile substantially suppresses the collapse and
extends the propagation distance of vortex bullets, their robustness is
ultimately limited by azimuthal instabilities, whose growth is enhanced for
values of $\ell $. These results demonstrate that engineered
refractive-index landscapes significantly enlarge the stability domain of
spatiotemporal vortex bullets and provide guidelines for supporting optical
bullets carrying large orbital angular momenta.
\end{abstract}

%-----------------------------------------------------------------------------
%\pacs{42.65.Tg, 05.45.Yv, 42.65.Jx}
%--------------------------------------------------------------------------------

\maketitle

\section{Introduction}

Spatiotemporal optical solitons \cite{malomed2005spatiotemporal}, commonly
referred to as light bullets, are multidimensional nonlinear wave packets
that keep their spatial and temporal self-trapping through the balance
between the diffraction, group-velocity dispersion (GVD), and nonlinear
self-focusing. Since the theoretical prediction by Silberberg in 1990 \cite%
{Silberberg1990collapse}, light bullets have been attracting much interest
due to their fundamental properties, as well as potential applications to
optical telecommunications, laser physics, propagation of ultra-strong light
pulses, and other fields \cite{kivshar2003optical}. Nowadays it 
is a field of intense research concerning the production of such 
bullets~\cite{Bock2025} as well as their existence and stability in a number of 
systems, as crystals with impurities~\cite{Sazonov2025}, optical 
fibers~\cite{Bejot2025}, lattices and fiber arrays~\cite{Chen2023,Aceves2015} 
or other types of designed media~\cite{Wang2025,Han2026}.  The creation 
and evolution of the spatiotemporal solitons are governed by the
multidimensional nonlinear Schr\"{o}dinger equation (NLSE)\cite%
{fibichnonlinear}, which provide the basis of the fundamental model for the
propagation of slowly varying envelopes of light pulses and beams.

It is well known that the optical bullets are unstable in self--focusing
materials with anomalous GVD~\cite{bespalov1966,sulem2007nonlinear}: above a
certain threshold energy the (2+1)-dimensional [(2+1)D] bullet undergoes
catastrophic collapse (blow-up) after a finite propagation distance,
while (3+1)D bullets are still more vulnerable to the collapse instability,
as the respective threshold~energy is
zero~\cite{edmundson1997unstable,fibichnonlinear}.
To prevent the blow-up and achieve stable propagation of nonlinear optical
wave packets in two or three spatial dimensions, various mechanisms have
been explored~\cite{Malomed2022}, including
the incorporation of higher--order terms in the nonlinear refractive index,
such as saturable~\cite{akhmediev1992modulation,edmundson1992} or
cubic--quintic nonlinearity~\cite{michinel2002liquid,michinel2006turning},
as well as nonlocality~\cite%
{krolikowski2001modulational,bang2002collapse,mihalache2006three,paredes2020optics}
and external potentials\cite{baizakov2003multidimensional}, among other
effects. In particular, optical bullets in materials with waveguiding
linear-index profiles can be stabilized through the trapping
provided by the effective potential representing the guiding profile \cite%
{mihalache2005stable,diebel2016controlled,akozbek1998optical},
including ingenious systems like moir\'e lattices~\cite{Kartashov2022}
or twisted waveguide arrays~\cite{Milian2019}.

Endowing these nonlinear wave packets with a topological charge (alias
winding number), which represents a phase singularity in the wave
field~\cite{nye1974dislocations,coullet1989}, leads to the formation of vortex
light bullets~\cite{desyatnikov2005azimuthons,desyatnikov2005optical}
carrying the orbital angular momentum~\cite{quiroga1997,michinel2004square},
in addition to being spatiotemporally self-trapped\cite{salamin2018fields}.
However, the ring-shaped intensity distribution in the vortex bullets is highly
susceptible to the azimuthal modulational instability, causing the vortex to
break up into several fundamental (zero-vorticity) solitons. Despite their
theoretical appeal and promising properties, the experimental realization of
stable optical bullets ---and especially of vortex bullets--- is highly
challenging~\cite{hang2013ultraslow}. Precise control over the input beam
and the balance of competing nonlinearities are critical for achieving the
sustained propagation~\cite%
{porras2019upper,chen1997self,zhang2022stabilization}. In the case of
discrete vortex bullets, the possibililty of the existence of transient stable states
in waveguiding arrays was experimentally demonstrated \cite{Eilenberger2013}. The
possibility of producing stable bullets in 2D Su-Schrieffer-Heeger
lattices with disclination defects was also demonstrated recently \cite{Zhang2026}. 
When an external index-guiding profile is used, the possibility of producing stable
fundamental (zero-vorticity) optical bullets is well known~\cite%
{mihalache2005stable}, although it was only recently theoretically
demonstrated that such a profile can support stable optical vortices with
winding number $1$, that propagate stably over distances much larger than
the underlying diffraction length and survive mutual collisions \cite%
{Sanchez2026}. Conditions under which the stable propagation is possible
require that the index-guiding profile must be sufficiently strong, provided
that the propagation constant belongs to a specific range of values.

In the present paper we aim to produce stable vortex bullets with relatively
large values of the winding number. It is well known that increasing angular
momentum normally enhances the azimuthal instability, leading to
fragmentation of stationary vortex states into a set of fundamental
solitons, thus making the stabilization of multidimensional vortex solitons
one of long-standing challenges in nonlinear optics. In this study, we
address a system similar to that considered in Ref.~\cite{Sanchez2026},
which includes the focusing Kerr nonlinearity and anomalous GVD in the
presence of a Gaussian-shaped linear-index guiding pattern in the
transversal plane.

The adopted model is introduced in the next section. The results for the
bullet families and their stability are reported in section \ref%
{sec:solution_families}. The paper is concluded by section 4.

\section{The model}

\label{sec:model}

We start from the (3+1)D NLSE with the Kerr nonlinear term,

\begin{equation}
i\frac{\partial \tilde{\Psi}}{\partial \tilde{z}}+\frac{1}{2k_{0}}\tilde{%
\nabla}_{\perp }^{2}\tilde{\Psi}-\frac{\beta _{2}}{2}\frac{\partial ^{2}%
\tilde{\Psi}}{\partial \tilde{T}^{2}}+k_{0}\left[ \Delta nV(r)+n_{2}|\tilde{%
\Psi}|^{2}\right] \tilde{\Psi}=0.  \label{eq:dimensional_NLSE}
\end{equation}%
where $\tilde{z}$ is the propagation distance, and Laplacian $\tilde{\nabla}%
_{\perp }^{2}$ is the diffraction operator\ acting in the transversal
spatial plane, while $k_{0}=\omega /c$, where $\omega $ is the carrier
frequency and $c$ the speed of light in vacuum. Further, $\tilde{T}=t-z/v_{g}
$, which includes the conventional time $t$, is the temporal coordinate in
the reference frame traveling with the group velocity $v_{g}$, $\beta _{2}<0$
is the coefficient of the anomalous GVD, and $n_2 >0$ is the coefficient
of the focusing Kerr nonlinearity. The axially symmetric index profile is
represented by function $V(\tilde{r})$ with amplitude $\Delta n$. By means
of rescaling,%$\tilde r=(\tilde x^2+\tildey^2)^{1/2}$
\begin{eqnarray}
r=k_{0}\sqrt{2\,\Delta n}\,\tilde{r}, &&\qquad z=k_{0}\,\Delta n\,\tilde{z},
\notag \\
T=\left( -\frac{2k_{0}\Delta n}{\beta _{2}}\right) ^{1/2}\tilde{T}, &&\qquad
\Psi =\left( \frac{n_{2}}{\Delta n}\right) ^{1/2}\tilde{\Psi},
\label{eq:scale_eqs}
\end{eqnarray}%
Eq. (\ref{eq:dimensional_NLSE}) is cast in the scaled form:

\begin{equation}
i%
\frac{\partial \Psi }{\partial z}+\nabla _{\perp }^{2}\Psi +\frac{\partial
^{2}\Psi }{\partial T^{2}}+\left[ V(r)+|\Psi |^{2}\right] \Psi =0.
\label{eq:NLSE}
\end{equation}%
The shape of the waveguiding index\ is fixed as the Gaussian,

\begin{equation}
V(r)=\gamma \,\exp {(-q\,r^{2})},  \label{eq:index_profile}
\end{equation}%
with amplitude $\gamma $ and width $w=1/\sqrt{q}$.

It is relevant to mention that Eq. (\ref{eq:NLSE}) with $z$ replaced by $t$,
and $T$ replaced by $z$, is identical to the Gross-Pitaevskii equation for the
self-attractive atomic Bose-Einstein condensate. In that case, potential given
by Eq.~(\ref{eq:index_profile}), with $\gamma >0$ and $q$ large
enough, represent the \textquotedblleft cigar-shaped" trapping potential for
atoms~\cite{Dunjko2001}. In that case, the vortex optical bullets are tantamount\ to 3D
matter-wave solitons with embedded vorticity.

Stationary vortex states, with integer winding number (topological charge) $%
\ell \geq 0$ and propagation constant $\beta $, are sought for, in terms of
the cylindrical coordinates $\left( r,z,\phi \right) $, as

\begin{equation}
\Psi (r,z,\phi ;T)=\psi (r,T)e^{i\ell \phi }e^{i(\beta z-\omega t)},
\label{eq:stationary_state}
\end{equation}%
where $\psi (r,T)$ is a real function, determined by the following equation
in the $\left( r,T\right) $ plane:%
\begin{equation}
-\beta \psi +\frac{\partial ^{2}\psi }{\partial T^{2}}+\frac{\partial
^{2}\psi }{\partial r^{2}}+\frac{1}{r}\frac{\partial \psi }{\partial r}-%
\frac{\ell ^{2}}{r^{2}}\psi +[V(r)+\psi ^{2}]\psi =0.
\label{eq:beta_equation}
\end{equation}%
Solutions of Eq. \ref{eq:beta_equation} are characterized by their
propagation constant $\beta $ and the value of the optical energy (alias
norm), $N$,

\begin{equation}
N=2\pi \int_{-\infty }^{\infty }\int_{0}^{\infty }|\psi |^{2}rdr\,dT,
\label{eq:energy}
\end{equation}%
which is a dynamical invariant of Eq. (\ref{eq:NLSE}). Below, we find
solutions of Eq. (\ref{eq:beta_equation}) and identify their stability for
increasing values of vorticity $\ell $, in the parameter plane of $\left(
\gamma ,q\right) $.

\section{Solution families and stability}

\label{sec:solution_families}

For numerical solution of Eq.(\ref{eq:beta_equation}), variables $(r,T)$
and real function $\psi $ were discretized using the finite-differences
scheme \cite{salgueiro2007computation}, and the resulting algebraic problem
was solved by iterations, by dint of the globally convergent Newton method.
Localized solutions are subject to conditions $\psi \rightarrow 0$ as $%
r\rightarrow \infty $ and $T\rightarrow \pm \infty $. At $r\rightarrow 0$,
the asymptotic form is obvious, $\psi \rightarrow r^{\ell }$.

\begin{figure}%%% 1
\par
\begin{center}
\includegraphics[width=\columnwidth]{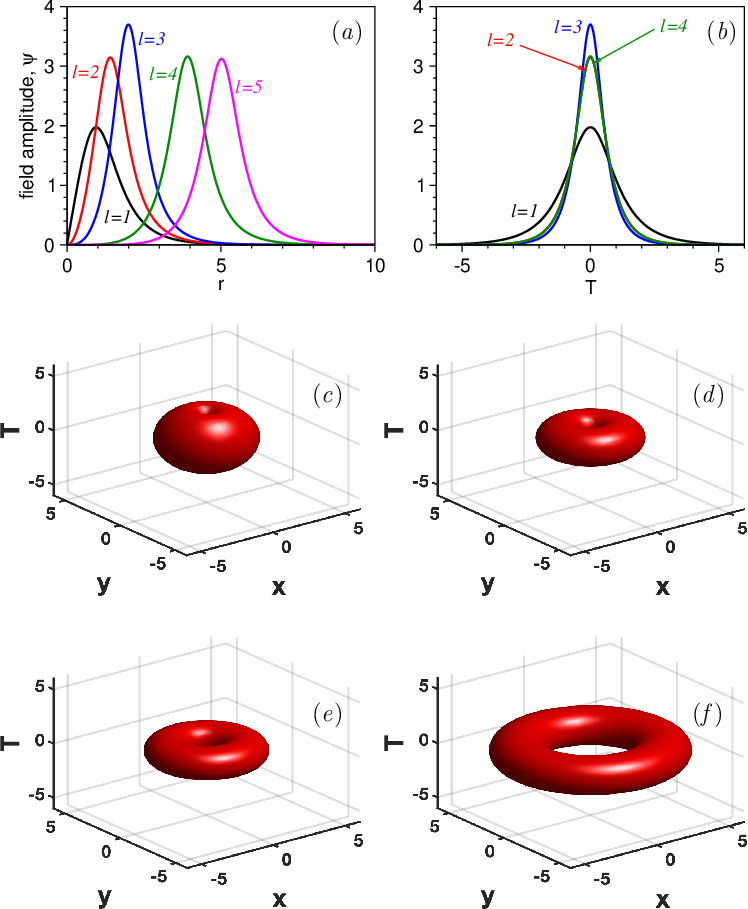}
\end{center}
\caption{Examples of optical bullets with embedded vorticity ranging
from $\ell=1$ to $4$, for $\gamma=4$, $q=1/2$ and $\beta=1.0$. (a) Radial
spatial profiles in cross-section $T=0$ (to which the maximum amplitude
belongs). (b) Temporal profiles in cross-section $r=r_{\max }$, where the
maximum amplitude is attained. (c-f) 3D toroidal shapes of vortex bullets
with winding numbers $\ell =1,2,3$ and $4$, respectively, represented by
means of isosurfaces $\protect\psi =\mathrm{0.25}.$}
\label{fig:bullet_examples}
\end{figure}

Examples of the solutions for vorticity values from $\ell =1$ to $4$ are
presented in Fig.~\ref{fig:bullet_examples}. In all cases prameter values are
fixed to $\gamma=4$, $q=1/2$ and $\beta=1.0$. Field profiles in the radial
and temporal directions are plotted in panels (a) and (b), while 3D toroidal
shapes of the vortex bullets are displayed by means of isosurfaces of $\psi$
in panels (c-f). A natural feature is the steep increase of the spatial width of
the central hole as $\ell$ increases. Also, the amplitude increases with $\ell$,
attaining a maximum at $\ell =3$, slightly decreasing and further saturating
at larger values of $\ell$. The temporal width in Fig.~\ref{fig:bullet_examples}(b)
initially decreases with the increase of $\ell $ up to $3$, and then
saturates.

\begin{figure}
%%%% 2
\par
\begin{center}
\includegraphics[width=\columnwidth]{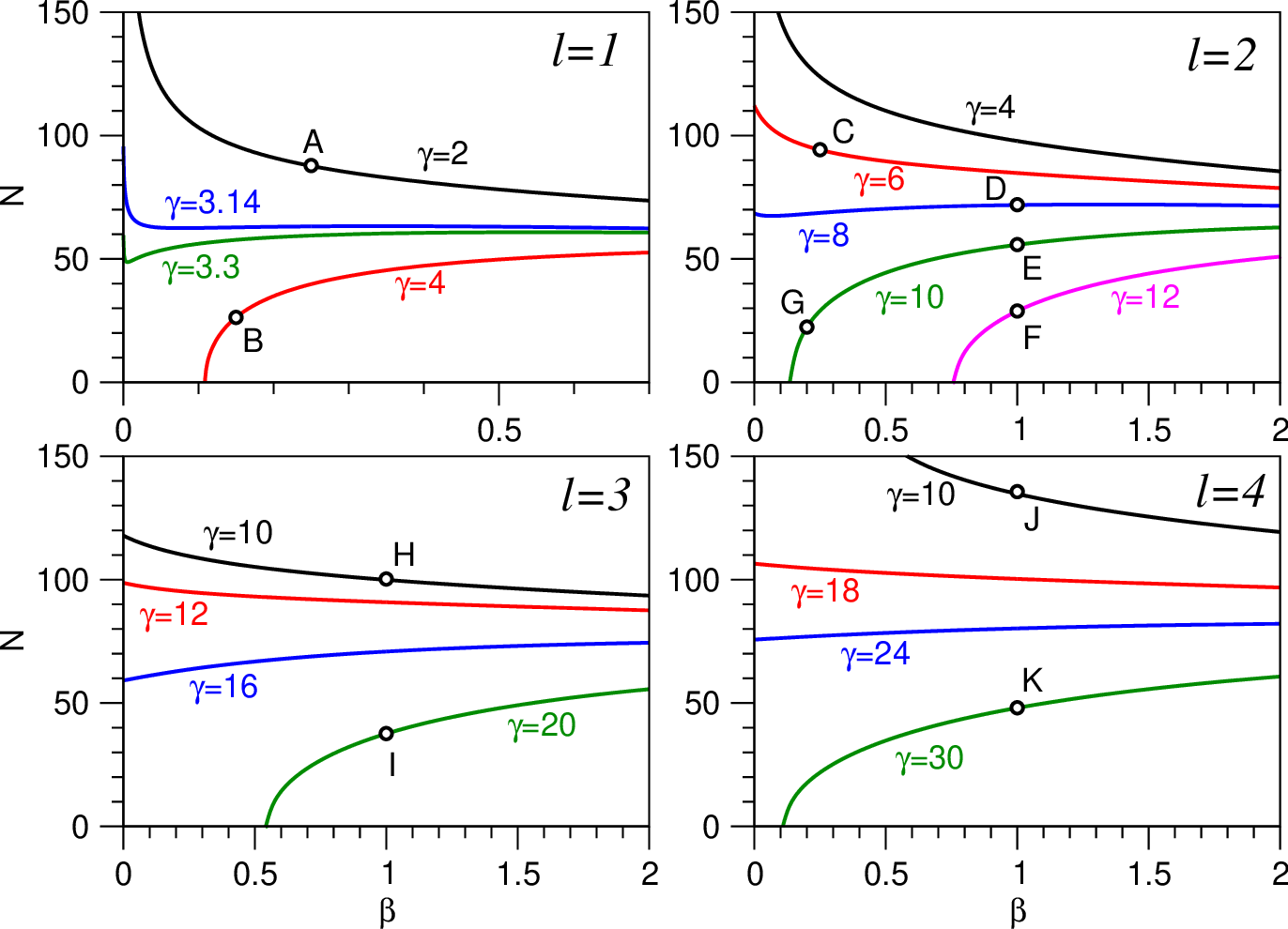}
\end{center}
\caption{The norm of vortex bullets vs. the propagation constant for
different values of the guiding-profile's strength $\protect\gamma $, fixed
width parameter $q=1/2$, see Eq. (\protect\ref{eq:index_profile}), and four
values of the winding number, from $\ell =1$ to $4$. Points marked
with letters correspond to the modes whose propagation is displayed in Fig.~\ref{fig:prop_q05} (the letters
match in both figures).}
\label{fig:Nvsbeta_q05}
\end{figure}

Families of the vortex-bullet solutions with vorticities $\ell =1,2,3$ and $4
$ are represented in Fig. \ref{fig:Nvsbeta_q05} by means of dependences $%
N(\beta )$, for $q=1/2$ and different values of $\gamma $ [see Eq. (\ref%
{eq:index_profile})]. These curves identify families meeting the celebrated
Vakhitov-Kolokolov (VK) criterion, which is the necessary (although not
sufficient) stability criterion for
solitons~\cite{Vakhitov1973,fibichnonlinear,Sanchez2026}.
It is observed that larger values of $\gamma$ (stronger guiding profiles)
are required to secure the validity of the VK criterion for the families
belonging to higher values of $\ell$.

%{\LARGE [It will be quite relevant to add a panel with }${\LARGE \ell =0}$%
%{\LARGE \ to Fig. 2, for the comparison with }${\LARGE \ell \geq 1}${\LARGE %
%.]}

In the case of the weak guiding channel, which corresponds to small
values of $\gamma$ in Eq.~(\ref{eq:index_profile}), the $N(\beta )$ %
dependence is close to that for strongly unstable 3D solitons in
free space, which is determined by the obvious scaling relation, $N\sim
\beta^{-1/2}$. It explains the divergence of $N(\beta )$ at
$\beta \rightarrow 0$ for ($\gamma =2, \ell=1$), ($\gamma =4, \ell=2$), and
($\gamma=10, \ell=4$) in Fig.~\ref{fig:Nvsbeta_q05} (in this limit, the
solitons are very broad, with the size scaling as $1/\sqrt{\beta}$, while
their amplitude is small and scales as $\sqrt{\beta}$). On the other hand,
the curves $N(\beta)$ which originate, at $N=0$,
from eigenvalues $\beta_{\mathrm{EV}}>0$ for larger $\gamma$ (e.g.,
$\gamma =4$ for $\ell =1$ and $\gamma =30$ for $\ell =4$) bifurcate from the
respective eigenmodes which exist at $\beta=\beta_{\mathrm{EV}}$
in the linearized version of Eq.~(\ref{eq:beta_equation}). Intermediate
values of $\gamma$, at which the bullet branches originate, at $\beta =0$,
from finite values $N>0$ in Fig.~\ref{fig:Nvsbeta_q05} (e.g., $\gamma =6$
and $8$ at $\ell=2$) correspond to 3D solitons, at $\beta =0$, with
slowly decaying tails.

\begin{figure}
%%%% 7
\par
\begin{center}
\includegraphics[width=\columnwidth]{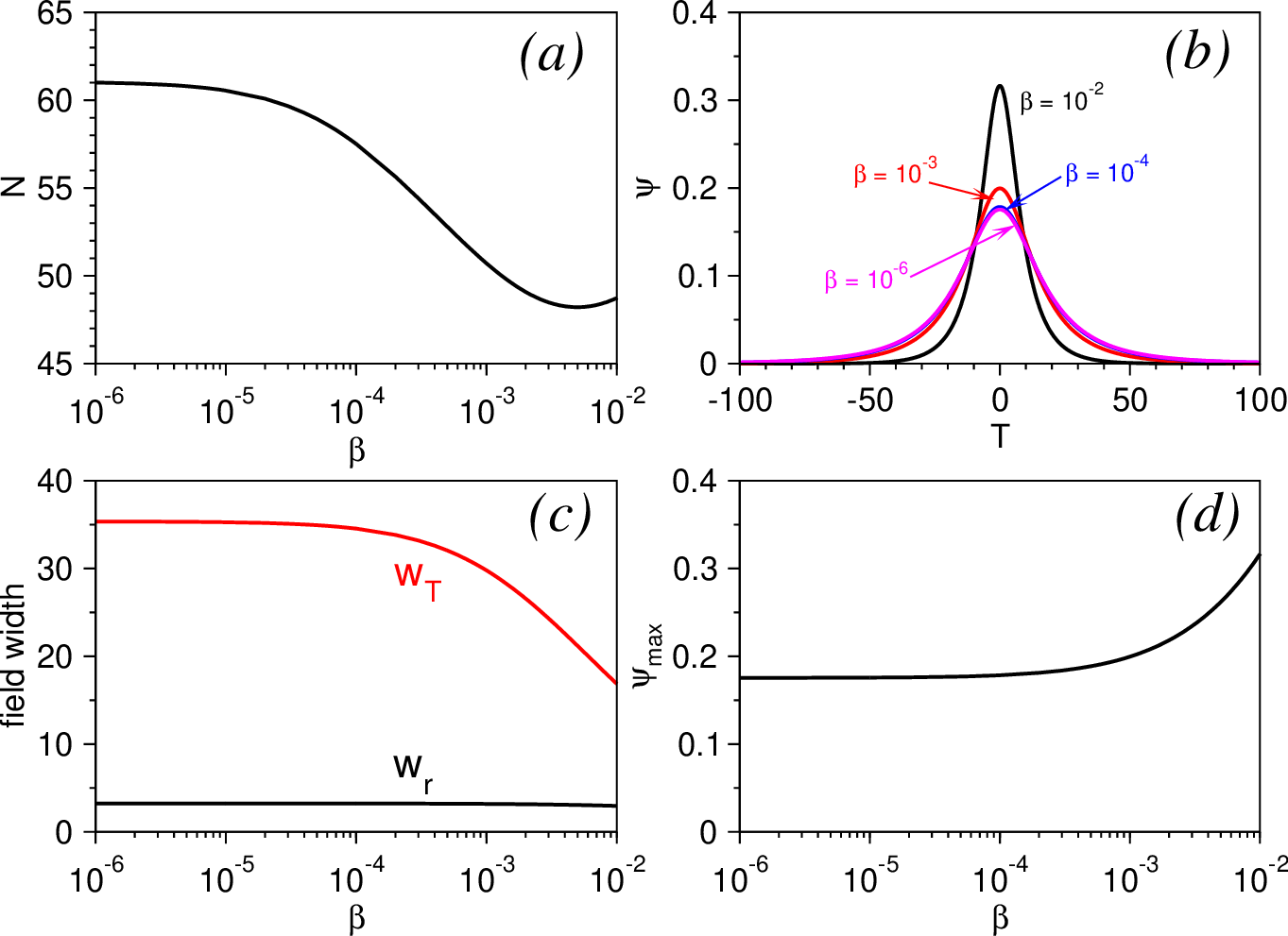}
\end{center}
\caption{(a) The norm vs. low values of the propagation constant for the family of
solutions with $\ell=1$, $\gamma=3.3$, $q=0.5$. (b) Temporal field profiles in the cross-section $r=r_{max}$,
corresponding to the field maximum. (c) Widths vs. $\beta$ of the field in the spatial ($w_r$) and temporal ($w_T$)
directions in the cross-sections corresponding to the field maximum. (d) The maximum field amplitude vs. $\beta$.}
\label{fig:low_beta_study}
\end{figure}

This intermediate behavior, i.e., the family of nonlinear
solutions originating at a finite value of $N$, is quite unusual and is an effect of the
waveguide combined with the nonlinearity as none of these factors can separately provide
an explanation for it. To check that the effect is not an artifact produced by a coarse step in varying
the value of $\beta$, as it approaches zero, or by a too narrow calculation domain, we have carried out
a careful analysis of the solutions at values of $\beta$ close to zero. The region of $\beta\to 0$
was explored in several stages, scanning values of $\beta$ on different scales. The results are summarized in
Fig.~\ref{fig:low_beta_study}.

Figure \ref{fig:low_beta_study}(a) shows the region of small $\beta$ on the logarithmic scale,
corresponding to the solution family with $\ell=1$, $\gamma=3.3$. Different examples
are presented in Fig.~\ref{fig:low_beta_study}(b), showing tails that decay to zero in the calculation domain,
which extends to the interval of $[-300,300]$ in both $r$ and $T$ (though a smaller domain is shown in the figure
for a better clarity), to secure a negligible effect of the boundaries on the solution. A fine mesh
of $3000\times 3000$ points in both coordinates was used for the numerical computations. This mesh was later made
finer, including $5000\times 5000$ and even $7000\times 7000$ points for the sake of comparison, to make it
sure that the discretization is suitable to produce correct results. In Fig.~\ref{fig:low_beta_study}(c) the field
size, calculated as the full width at half maximum (FWHM), is shown for both spatial and temporal profiles.  The
width initially increases, as $\beta$ varies from larger values toward $\beta=0$ [cf. the examples displayed
in Fig.~\ref{fig:bullet_examples}(b)]. This happens mainly for the temporal profile, as the
waveguide helps to keep the spatial width at lower values.  Further, the width saturates and converges to a finite
value at $\beta\to 0$. Similarly, the maximum amplitude decreases, converging to a nonzero value
[Fig.~\ref{fig:low_beta_study}(d)].

% The following text can be put if the figure is suppressed
%The results reveal that the field width initially increases fast, up to a point where it saturates to finally converge to a finite value.  Similarly, maximum amplitude decreases to converge to a fixed value different from zero. This means that also norm (energy) N converges to a finite value. In order to check the validity of this solutions we have calculated the solutions maintaining a wide enough domain and a large enough number of samples. At the point of convergence, the field tails are well kept in the interval $[-300,300]$, both for $r$ and $T$, and a mesh of $3000\times 3000$ points in both coordinates was used for the numerical calculus. This mesh was later raised to $5000\times 5000$ and even to $7000\times 7000$ for comparison, in order to assure that the discretization is fine enough to produce correct results.}

Fig.~\ref{fig:Nvsbeta_q05} suggests that there might exist a special value
of $\gamma $ (close to
$\gamma =8$ at $\ell=2$, $\gamma=16$ at $\ell=3$, and to
$\gamma=21$ at $\ell=4$), which corresponds to a separatrix between
the VK-complying and non-complying branches, with $N(\beta )=\textrm{const.}$ %
Such a degenerate branch would represent a family of Townes
solitons~\cite{fibichnonlinear}. Being formally neutral with respect to the VK criterion,
Townes solitons are subject to a weakly growing instability, that eventually leads to the onset of
the collapse~\cite{fibichnonlinear}. In reality, detailed numerical results demonstrate that $N$ always decreases
at sufficiently large values of $\beta$. What is actually found is a
value of $\gamma$ marking the transition from monotonously-decreasing $N(\beta)$ curves to ones
with a maximum of $N(\beta)$, that separates regions with $dN/d\beta <0$ and $dN/d\beta >0$.
This value, which is, e.g., $\gamma\sim 3.14$ for $\ell=1$ in
Fig.~\ref{fig:Nvsbeta_q05}, has an almost null slope for small values of $\beta$,
resembling a branch of Townes solitons, but it ultimately rises to a higher finite
value of $N$ at $\beta\to 0$. Further increasing $\gamma$, a value is
attained at which the bifurcation from the linear mode commences.

%{\LARGE [The above highlighted text is added by me. The statements
%concerning the divergence of N in the limit of }${\LARGE \beta \rightarrow 0}
%${\LARGE \ for small }${\LARGE \gamma }${\LARGE , and the bifurcation of the
%}${\LARGE N(\beta )}${\LARGE \ branches from the linear eigenmodes at }$%
%{\LARGE N=0}${\LARGE \ for small }${\LARGE \gamma }${\LARGE \ are definitely
%correct. However, the statement that \textquotedblleft }{\large Intermediate
%values of }$\gamma ${\large , at which the bullet branches originate, at }$%
%\beta =0${\large , from finite values }$N>0${\large \ in Fig. \ref%
%{fig:Nvsbeta_q05} (e.g., }$\gamma =6${\large \ and }$8${\large \ at }$\ell =2
%${\large ) correspond to 3D solitons, at }$\beta =0${\large , with slowly
%decaying tails}{\LARGE ." is my wild guess. Is it possible to check it and
%illustrate by an example of such a soliton with finite }${\LARGE N}${\LARGE %
%\ at }${\LARGE \beta }${\LARGE \ close to }${\LARGE 0?}${\LARGE \ The
%statement concerning the existence of the degenerate branch, with }${\LARGE %
%N(\beta )\equiv }${\LARGE const, is another wild guess. Is it possible to
%check it too?]}

% My response: The statement concerning the slowly decaying tails is correct. The existence of the degenerate branch with a N(\beta)=const is not as all branches decrease for \beta large enough. At most we can say that there is a value of \gamma for which N(\beta) remains approximately constant for small enough values of \beta.

The VK criterion is not sufficient for the stability and, additionally, the fact that the
optical bullets are endowed with angular momentum implies the
possibility of azimuthal instabilities, which the VK criterion cannot address. It is then
necessary to check the stability by means of numerical simulations of the propagation.
For $\ell =1$ it was already found that the stability can be maintained for
the propagation distances larger than the respective Rayleigh (diffraction) length, $%
z_{R}=k_{0}\,w_0^{2}/2$, where $w_{0}$ is the spatial width of the optical
bullet. For instance, the profiles displayed in Fig.~\ref{fig:bullet_examples} have $%
w_{0}<10$, with $z_{R}\sim 50$. Simulations were carried out
applying the beam propagation method (BPM) to the 3D equation
[Eq.~(\ref{eq:NLSE})] and using different vortex bullets as initial conditions. For this purpose, a
finite-difference scheme was used to discretize the transverse variables
$(x,y)$ and the temporal one ($T$). The equation kernel was applied in two
steps, following the usual splitting operator techniques~\cite{Press1992}, to obtain the
propagated field along the step $dz$.  First, the Laplace operator is applied, using an
alternating-direction implicit (ADI) method, using transparent boundary conditions at all domain
boundaries~\cite{Hadley1992}. As this algorithm is not unconditionally
numerically-stable in three dimensions, an improved version of the ADI algorithm was
used~\cite{Chang1991}. Second, the potential $V(r)$ and nonlinear term were
applied, using the fourth-order Runge-Kutta algorithm~\cite{Press1992}.  The propagation
simulations in this work where performed using a mesh of $300\times 300\times 300$ points
and a step size $dz=0.002$.

\begin{figure}
%%%% 7
\par
\begin{center}
\includegraphics[width=\columnwidth]{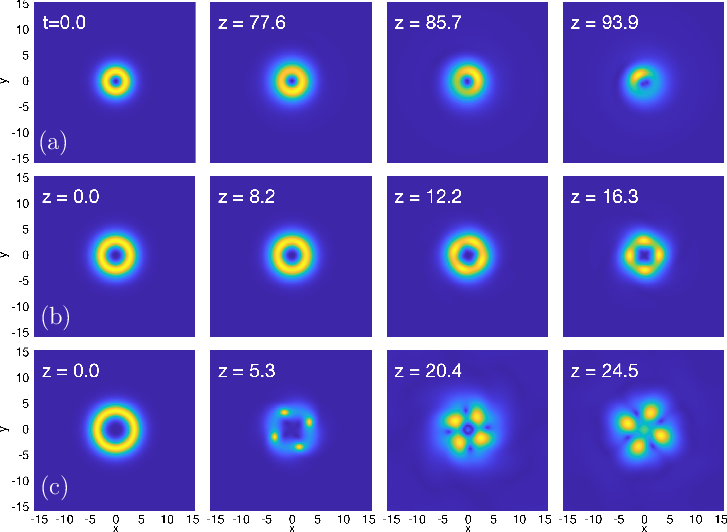}
\end{center}
\caption{The development of the splitting (azimuthal) instability for the
toroidal vortex bullets with winding numbers $\ell =2$, $3$, and $4$. In all
cases, the guiding-profile strength is $\protect\gamma =4$. (a) $\ell =2$, $%
q=0.16$, $\protect\beta =1.0$; (b) $\ell =3$, $q=0.08$, $\protect\beta =1.0$;
(c) $\ell =4$, $q=0.04$, $\protect\beta =1.5$. The panels show
the plane $(x,y)$ with $T=0$, where the field maximum is located.}
\label{fig:decay_patterns}
\end{figure}

\begin{figure}
%%%% 7
\par
\begin{center}
\includegraphics[width=\columnwidth]{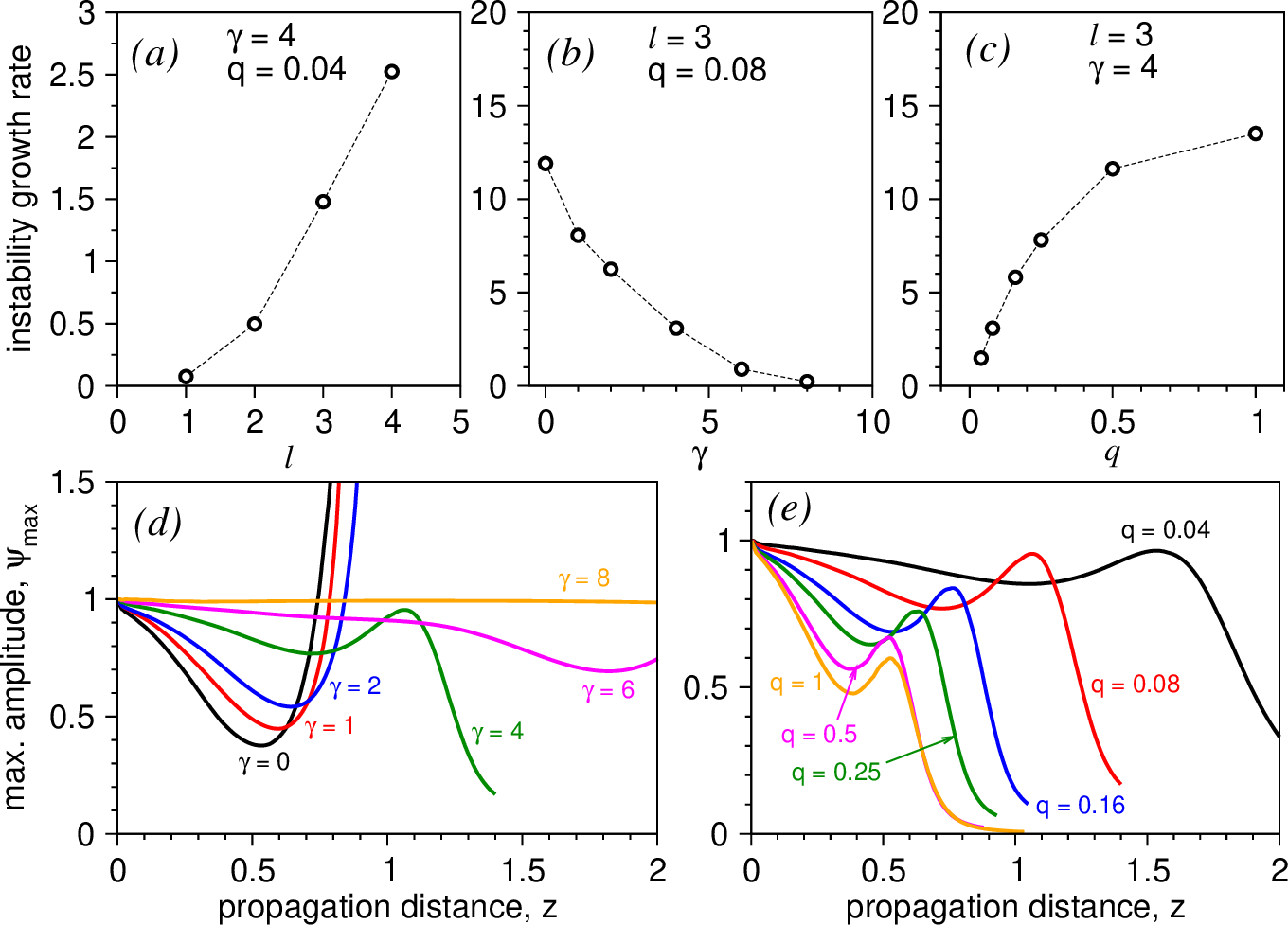}
\end{center}
\caption{(a-c)~The instability growth rate as a function of different system's parameters: 
topological charge (a), index strength (b), and the profile's width $w=1/\sqrt{q}$ (c), with fixed values of 
other parameters, as indicated in each panel. The propagation constant is $\beta=3$ in all cases. Dashed thin lines 
connecting the points are plotted as a guide for the eye. (d)~The maximum field amplitude plotted against the
propagation distance for the points shown in panel (b). (e)~The same for the points represented in panel (c).}
\label{fig:azimuthal_inestability}
\end{figure}

Typical examples of the instability development of the toroidal vortex
bullets with winding numbers $\ell =2,3$ and $4$, initiated by the
above-mentioned azimuthal instability, are presented in Fig.~%
\ref{fig:decay_patterns}, in the form of 2D images representing the plane of the
vortex ring. It is seen that the instability which splits the
axially symmetric structures develops much faster for $\ell =3$ and $4$ than
for $\ell =2$. To carry out a more quantitative study of this
instability, we have performed the simulations for optical bullets with different values of the parameters,
namely the charge $\ell$, index strength $\gamma$, and index-profile width $q$, maintaining constant values
of other two parameters in all the three cases. The propagation constant was fixed as
$\beta=3$, as for such a value there exists a solution (bullet) for a wide range of parameters, and also
because the azimuthal instability sets in after a relatively short propagation distance. The results are
summarized in Fig.~\ref{fig:azimuthal_inestability}. The instability growth rate represented in panels (a-c)
is evaluated as the inverse of the distance at with the vortex breaks, taking, as a breaking criterion,
the position at which the maximum amplitude of the field changes by 10\% against the
initial one.

In Fig.~\ref{fig:azimuthal_inestability}(a) the instability growth rate is plotted vs. the winding number, 
confirming the fast growth of the instability as $\ell$ increases. On the other hand, at
a particular value of the winding number, the instability growth rate quickly decreases as the
waveguide strength $\gamma$ increases, as shown in Fig.~\ref{fig:azimuthal_inestability}(b),
approaching zero at large values of $\gamma$, implying slow development of the instability development, 
which can lead to practical suppression of the instability in some cases [see also ~\ref{fig:azimuthal_inestability}(d), where the maximum amplitude of the field is plotted against propagation constant].
%The maximum amplitude of the field is plotted in Fig.~\ref{fig:aziest}(d), showing that a value of $\gamma$
%large enough (see the case $\gamma=8$ for which vortex splitting takes place at  $z\sim 4.5$
Finally, in Fig.~\ref{fig:azimuthal_inestability}(c) the instability growth rate is plotted against
the index-profile's width. The maximum field amplitude vs. the propagation distance for the corresponding points is shown in Fig.~\ref{fig:azimuthal_inestability}(e). In this case, the increase of the profile's width (i.e., decrease of $q$) indeed delays the onset of the azimuthal instability. As a summary, we conclude that stronger guiding, provided by the deeper index-modulation profile, helps 
to suppress the spontaneous destruction of the vortex ring. Lower values of the propagation constant $\beta$ also produce much weaker azimuthal instability (at least for relatively small winding numbers $\ell$).

\begin{figure}%%%% 3
\par
\begin{center}
\includegraphics[width=\columnwidth]{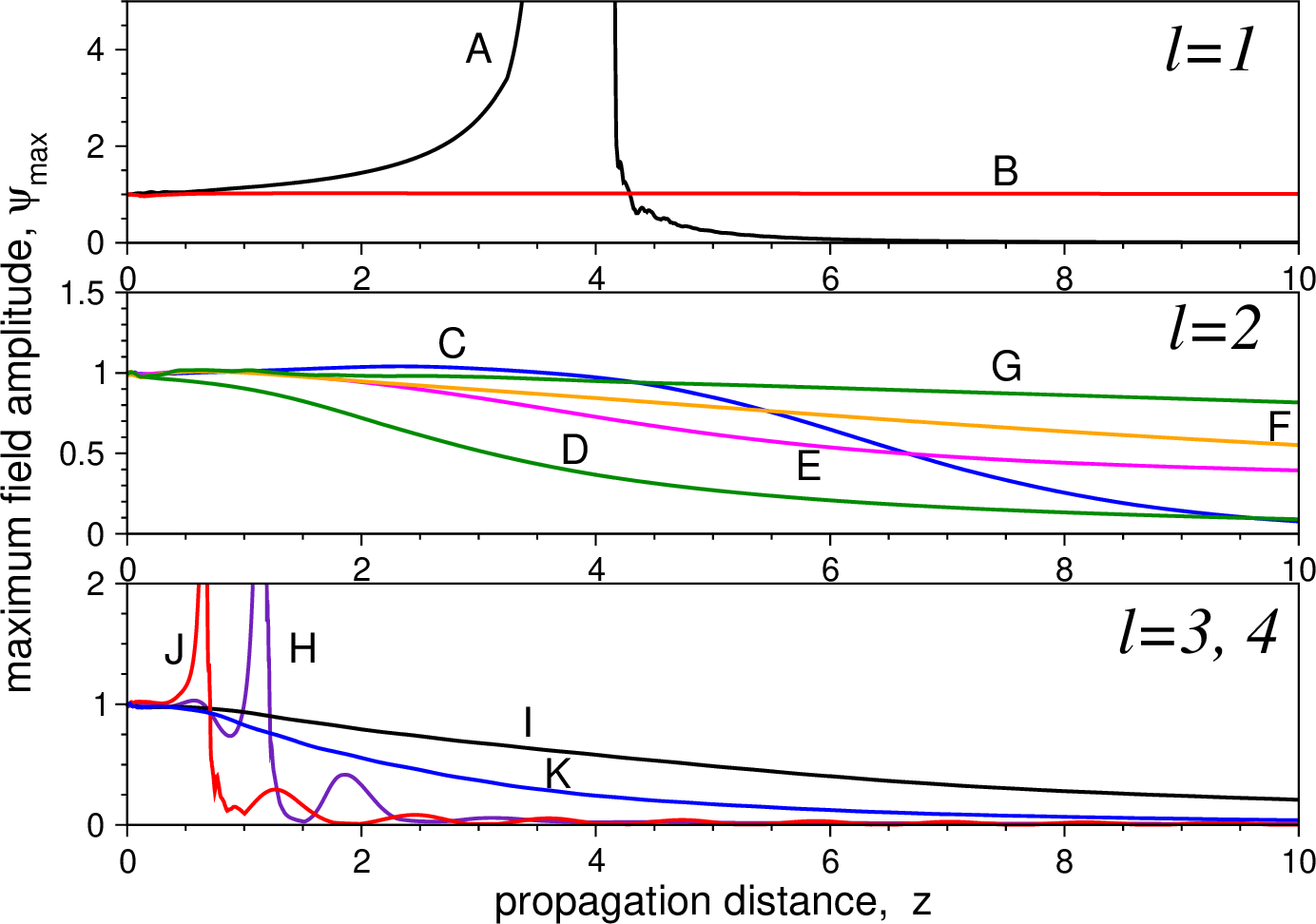}
\end{center}
\caption{The amplitude of the field in numerical simulations of the
propagating vortex bullet, scaled by the input value at $z=0$, vs. the
propagation distance for different stationary states marked in Fig.~\protect
\ref{fig:Nvsbeta_q05}. Letters label curves corresponding to the
modes marked in Fig.~\ref{fig:Nvsbeta_q05}.}
\label{fig:prop_q05}
\end{figure}

In Fig.~\ref{fig:prop_q05} we display systematic results, obtained by
plotting the evolution of the field's amplitude vs. the propagation distance
$z$. The states belonging to the VK-unstable branches, such as those
labelled A, J and H (for $\ell =1, 3$ and $4$, respectively) quickly suffer
the collapse, attaining high values of the amplitude (which is followed by rapid
decay, which may be an effect depending on the numerical scheme). On the
other hand, the VK-stable states (belonging to the branches with $dN/d\beta
>0$) do not collapse and tend to slowly spread out in the temporal
direction, while their spatial size remains confined by the guiding profile.
The spreading rate depends on the guiding strength $\gamma$, propagation
constant $\beta$, and mainly on the vorticity. In particular, for $\ell=1$
the amplitude maintains a nearly constant value for a long propagation
distance (curve B in Fig.~\ref{fig:prop_q05}). The vortex bullets
decay essentially faster for larger values of $\ell$ (compare
curve B for $\ell=1$ to D, E, F for $\ell=2$, and curves I and K for $\ell
=3$ and $4$, respectively).

%{\LARGE [It is relevant to add, to Fig. 3, a couple of examples of the
%evolution of }${\LARGE \psi }_{\max }${\LARGE \ of the state with }${\LARGE %
%\ell =0}${\LARGE .]}

For a fixed vorticity, the effective stability range increases with the
increase of the guiding-profile strength $\gamma $ (see, e.g., curves D, E
and F in Fig.~\ref{fig:prop_q05} for $\ell =2$ and $\gamma =8$, $10$ and $12$%
, respectively), which is similar to the effect securing the VK criterion by
larger values of $\gamma $ in Fig. \ref{fig:Nvsbeta_q05}. As one can see
in Fig.~\ref{fig:Nvsbeta_q05}, $\gamma=4$ is sufficient for $\ell=1$,
while it is necessary to have $\gamma \gtrsim 8$ for $\ell=2$, $\gamma
\gtrsim 16$ for $\ell=3$, and $\gamma \gtrsim 24$ for $\ell=4$. On the other
hand, the effective stability does not strongly depend on the propagation
constant $\beta$, although the decay is slightly faster for larger $\beta$
(compare, for instance, curves G and E).

The strong instability of the vortex bullets for vorticities $\ell\geq 2$
suggests one to seek additional stabilization scenarios in
this case. In particular, it is interesting to consider wider guiding
profiles, as it is then necessary to apply a higher power for balancing
the spatial diffraction by the Kerr nonlinearity; in turn, the stronger
self-attraction may enhance the confinement in the temporal direction.
This fact is essential, as the bullet's spreading occurs in such temporal direction.

\begin{figure}%%%% 4
\par
\begin{center}
\includegraphics[width=\columnwidth]{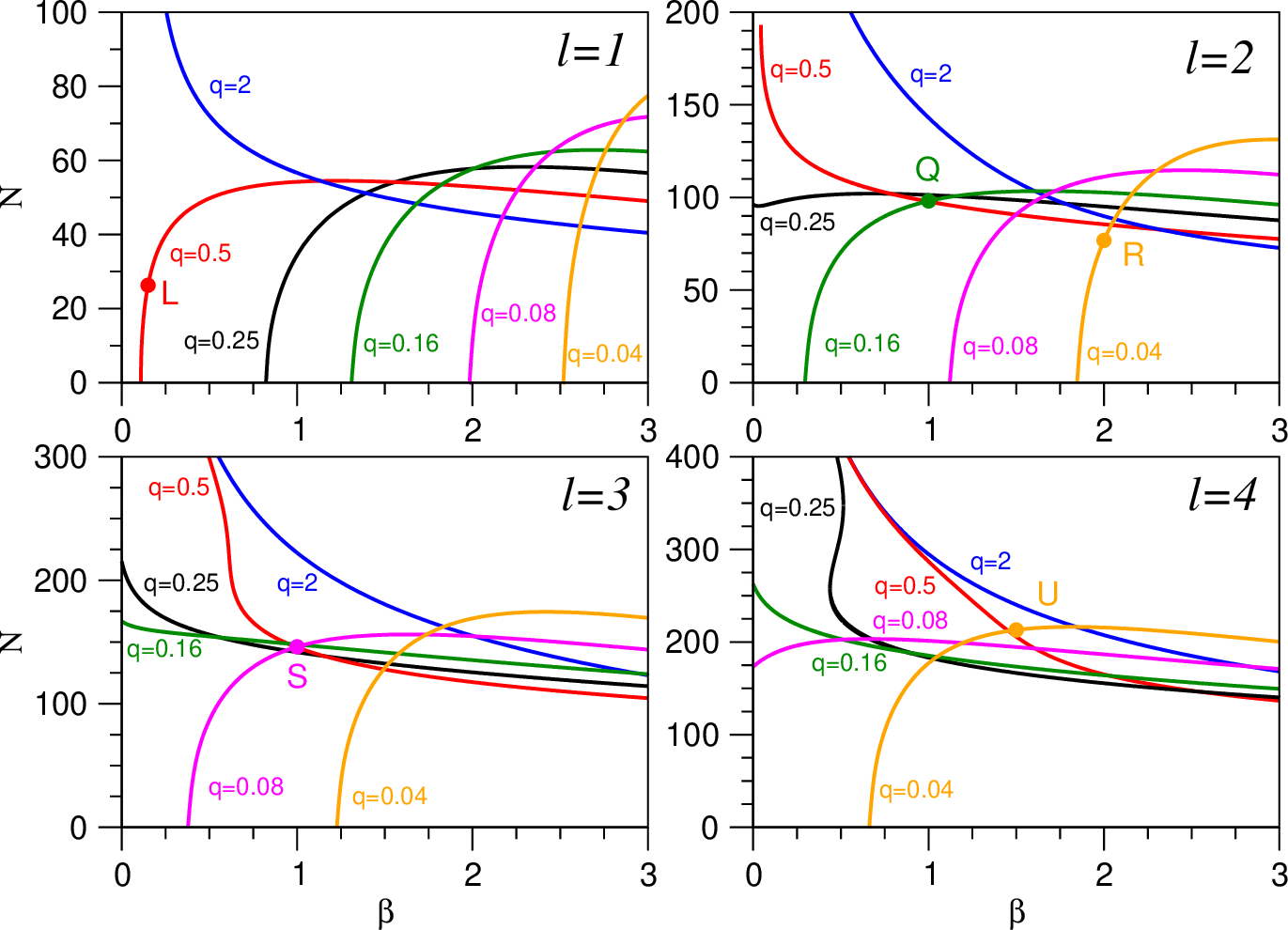}
\end{center}
\caption{The norm of the vortex bullets vs. the propagation constant for
different values of the width parameter $q$ and a fixed strength of the
guiding profile, $\protect\gamma =4$, in Eq.~(\protect\ref{eq:index_profile}%
). Each panel corresponds to a particular value of the winding number $\ell$. 
Points labeled by letters correspond to the propagating modes
shown in Fig.~\ref{fig:prop_gm4}.}
\label{fig:Nvsbeta_gm4}
\end{figure}

The effect of the variation of the width parameter $q$ is presented in Fig.~%
\ref{fig:Nvsbeta_gm4}, for a fixed guiding strength, $\gamma =4$, and the
same vorticities as above, from $\ell =1$ to $4$. As expected, the VK
criterion holds, making the vortex bullets more robust for smaller values
of $q$, wider guiding profiles, while larger $q$ strongly destabilize the
modes by making $dN/d\beta <0$. The interpretation of the solution branches
with $N(\beta )$ which diverge at $\beta \rightarrow 0$, or originate at $N=0
$ from the linear modes with $\beta =\beta _{\mathrm{EV}}>0$, and terminate,
at $\beta =0$, at finite values of $N$ is the same as outlined above in the
connection to Fig.\ref{fig:Nvsbeta_q05}.

%{\LARGE [Is it possible to add plots }${\LARGE N(\beta )\equiv }${\LARGE %
%const in Fig. 4, for specific values of }${\LARGE q}$ {\LARGE corresponding
%to the Townes soliton subfamily? I have added the mention of the Townes
%solitons to the Abstract and Conclusion.]}

\begin{figure}%%%% 5
\par
\begin{center}
\includegraphics[width=\columnwidth]{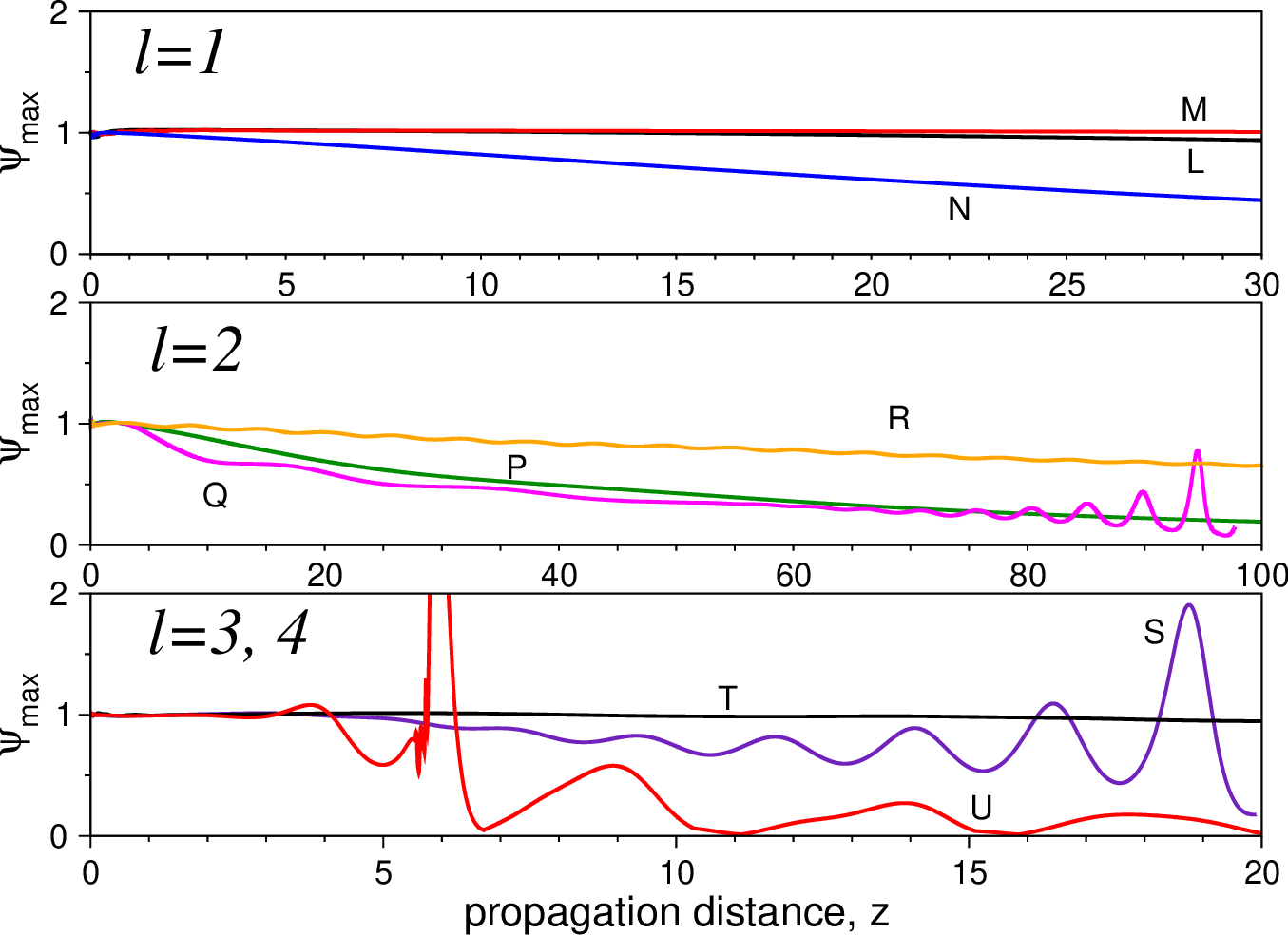}
\end{center}
\caption{The same as in Fig. \protect\ref{fig:prop_q05}, but for the vortex
bullets which correspond to dots marked by letters L, Q, R, S, and U in
Fig.~\protect\ref{fig:Nvsbeta_gm4}, and additional ones, \textit{viz}., M ($\ell=1$, %
$\protect\gamma =2$, $q=0.25$, $\protect\beta =0.08$), N ($\ell=1$,
$\protect\gamma =14$, $q=2$, $\protect\beta =0.1$), P ($\ell=2$, $
\protect\gamma=6$, $q=0.25$, $\protect\beta =0.6$), and T ($\ell=3$,
$\protect\gamma =4$, $q=0.01$, $\protect\beta =2.7$). }
\label{fig:prop_gm4}
\end{figure}

Fig.~\ref{fig:prop_gm4} displays results of the propagation-stability test
for representative examples of the vortex bullets, corresponding to dots L,
Q, R, S, U marked in Fig. \ref{fig:Nvsbeta_gm4}, and additional examples
defined in the caption to Fig. \ref{fig:prop_gm4}. It is seen that, at least
for $\ell =1$, it is possible to find an optimum value of $q$ which secures
the effectively stable propagation over a desired distance. For $\ell =2$ or
$3$, a reasonably large distance of the robust propagation, amounting to a
few Rayleigh lengths, can be provided too. An example is curve T, which
corresponds to small $q=0.01$, i.e., a wide guiding profile with width $%
w=1/\sqrt{q}=10$. The stable propagation is harder to be supported for
higher values of $\ell $, at which the above-mentioned azimuthal
modulational instability of the toroidal vortex bullets quickly kicks in. In
Fig.~\ref{fig:prop_gm4} this is shown, e.g., by curves S or U, while in the
above-mentioned quasi-stable case of curve T, the azimuthal instability only
commences at very large values of $z$.

\begin{figure}
%%%% 6
\par
\begin{center}
\includegraphics[width=\columnwidth]{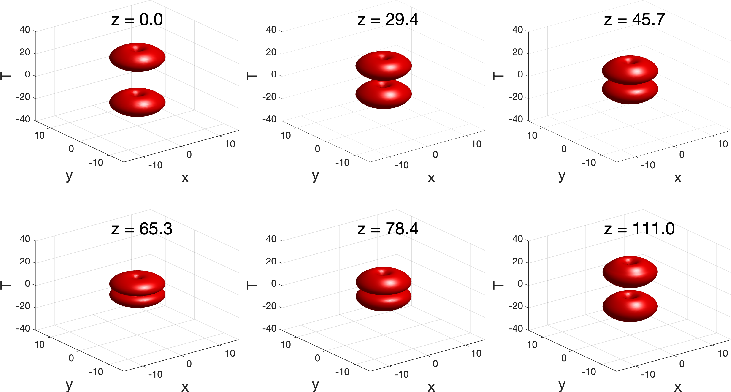}
\end{center}
\caption{The simulated elastic collision between two quasi-stable toroidal
vortex bullets with $\ell =2$, which correspond to symbol R in Figs.~\protect
\ref{fig:Nvsbeta_gm4} and \protect\ref{fig:prop_gm4}. The parameters are $%
\protect\gamma =4$, $q=0.04$, $\protect\beta =2$. }
\label{fig:collision_stable}
\end{figure}

Simulations of collisions between effectively stable vortex bullets in the
temporal direction demonstrate that the collisions are elastic, which is an
additional argument in favor of the robustness of these modes. As an
example, the collision between identical bullets with vorticity $\ell =2$,
which correspond to symbol R in Figs.~\ref{fig:Nvsbeta_gm4} and \ref%
{fig:prop_gm4}, is displayed in Fig.~\ref{fig:collision_stable}. In this
case, even though the slow decay of individual bullets has started prior to
the collision, the bullets perfectly persist and bounce back elastically.
The colliding bullets in the $T$ direction are generated introducing
a linear phase $\exp[\pm i\,(\sin\theta)\,T]$, which introduces an
angle $\theta$ between the propagation and time ($T$) directions. For this particular
simulation, the angle was chosen as $\theta=\pi/24$, which results in the
collision velocity $v_T=0.26$ along the $T$ direction. A similar result is
demonstrated by the colliding vortex bullets
with $\ell =3$, if their width parameter is reduced to $q=0.01$.

%{\LARGE [In the caption to Fig. \ref{fig:collision_stable}, it is necessary
%to indicated the (inverse) velocity of the collision. Maybe it is possible
%to add a result for the collision between the bullets with }$\ell =\pm 2$%
%{\LARGE ?].}

\begin{figure}
\par
\begin{center}
\includegraphics[width=\columnwidth]{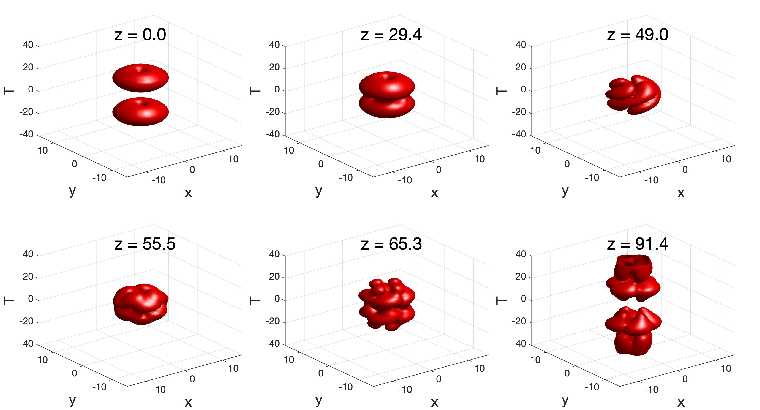}
\end{center}
\caption{The same simulation as in Fig.~\ref{fig:collision_stable} but for opposite
charged bullets ($\ell=\pm2$ respectively). Parameters: $%
\protect\gamma =4$, $q=0.04$, $\protect\beta =2$. }
\label{fig:collision_opcharge}
\end{figure}

In Fig.~\ref{fig:collision_opcharge} we present the same
simulation but for colliding bullets with opposite charges, $\ell=\pm 2$,
which implies zero total angular momentum (the collision with ``hidden vorticity").
As it was found in other
systems\cite{Salgueiro2016,Izdebskaya2012,Brtka2010}, outcomes of the collisions
may be completely different from the case of the collision between bullets with identical vorticities.
As seen in Fig.~\ref{fig:collision_opcharge}, collisions between bullets with opposite vorticities
are destructive.

\section{Conclusion}

To address the known problem of the stabilization of multidimensional
solitons with embedded vorticity, we have proposed the 3D nonlinear-Schr\"{o}%
dinger model, which includes the anomalous GVD, cubic self-attractive
nonlinearity, and a guiding index structure in the transverse plane, in the
form of an axisymmetric Gaussian. The same model applies to the
self-attractive atomic Bose-Einstein condensate. The numerical analysis has
demonstrated that the value of the vorticity ($\ell =1,2,3,4$) is the main
factor affecting the (in)stability of the 3D vortex bullets, along with the
strength and width of the guiding profile. Generally, the instability is
quickly enhanced with the increase of $\ell$. The propagation of the
bullets is more robust for the solutions satisfying the VK
(Vakhitov-Kolokolov) criterion. A degenerate family of the Townes solitons
is found at the boundary between VK-stable and unstable solutions. For a
given vorticity, it is possible to find an operational regime, with
sufficiently large strength and width of the guiding structure, which
provide effectively stable propagation of the bullets over a required
distance. Collisions between the effectively stable vortex bullets, in the
temporal direction, are fully elastic.

\section*{Acknowledgements}

This publication is part of the R\&D\&i project PID2023-146884NB-I00 funded
by Ministerio de Ciencia e Innovación. This work was also supported by grant
ED431B 2021/22 (Xunta de Galicia).

\bibliography{references01}

\begin{thebibliography}{52}%
\makeatletter
\providecommand \@ifxundefined [1]{%
 \@ifx{#1\undefined}
}%
\providecommand \@ifnum [1]{%
 \ifnum #1\expandafter \@firstoftwo
 \else \expandafter \@secondoftwo
 \fi
}%
\providecommand \@ifx [1]{%
 \ifx #1\expandafter \@firstoftwo
 \else \expandafter \@secondoftwo
 \fi
}%
\providecommand \natexlab [1]{#1}%
\providecommand \enquote  [1]{``#1''}%
\providecommand \bibnamefont  [1]{#1}%
\providecommand \bibfnamefont [1]{#1}%
\providecommand \citenamefont [1]{#1}%
\providecommand \href@noop [0]{\@secondoftwo}%
\providecommand \href [0]{\begingroup \@sanitize@url \@href}%
\providecommand \@href[1]{\@@startlink{#1}\@@href}%
\providecommand \@@href[1]{\endgroup#1\@@endlink}%
\providecommand \@sanitize@url [0]{\catcode `\\12\catcode `\$12\catcode
  `\&12\catcode `\#12\catcode `\^12\catcode `\_12\catcode `\%12\relax}%
\providecommand \@@startlink[1]{}%
\providecommand \@@endlink[0]{}%
\providecommand \url  [0]{\begingroup\@sanitize@url \@url }%
\providecommand \@url [1]{\endgroup\@href {#1}{\urlprefix }}%
\providecommand \urlprefix  [0]{URL }%
\providecommand \Eprint [0]{\href }%
\providecommand \doibase [0]{https://doi.org/}%
\providecommand \selectlanguage [0]{\@gobble}%
\providecommand \bibinfo  [0]{\@secondoftwo}%
\providecommand \bibfield  [0]{\@secondoftwo}%
\providecommand \translation [1]{[#1]}%
\providecommand \BibitemOpen [0]{}%
\providecommand \bibitemStop [0]{}%
\providecommand \bibitemNoStop [0]{.\EOS\space}%
\providecommand \EOS [0]{\spacefactor3000\relax}%
\providecommand \BibitemShut  [1]{\csname bibitem#1\endcsname}%
\let\auto@bib@innerbib\@empty
%</preamble>
\bibitem [{\citenamefont {Malomed}\ \emph {et~al.}(2005)\citenamefont
  {Malomed}, \citenamefont {Mihalache}, \citenamefont {Wise},\ and\
  \citenamefont {Torner}}]{malomed2005spatiotemporal}%
  \BibitemOpen
  \bibfield  {author} {\bibinfo {author} {\bibfnamefont {B.~A.}\ \bibnamefont
  {Malomed}}, \bibinfo {author} {\bibfnamefont {D.}~\bibnamefont {Mihalache}},
  \bibinfo {author} {\bibfnamefont {F.}~\bibnamefont {Wise}},\ and\ \bibinfo
  {author} {\bibfnamefont {L.}~\bibnamefont {Torner}},\ }\bibfield  {title}
  {\bibinfo {title} {Spatiotemporal optical solitons},\ }\href@noop {}
  {\bibfield  {journal} {\bibinfo  {journal} {Journal of Optics B: Quantum and
  Semiclassical Optics}\ }\textbf {\bibinfo {volume} {7}},\ \bibinfo {pages}
  {R53} (\bibinfo {year} {2005})}\BibitemShut {NoStop}%
\bibitem [{\citenamefont {Silberberg}(1990)}]{Silberberg1990collapse}%
  \BibitemOpen
  \bibfield  {author} {\bibinfo {author} {\bibfnamefont {Y.}~\bibnamefont
  {Silberberg}},\ }\bibfield  {title} {\bibinfo {title} {Collapse of optical
  pulses},\ }\href@noop {} {\bibfield  {journal} {\bibinfo  {journal} {Optics
  letters}\ }\textbf {\bibinfo {volume} {15}},\ \bibinfo {pages} {1282}
  (\bibinfo {year} {1990})}\BibitemShut {NoStop}%
\bibitem [{\citenamefont {Kivshar}\ and\ \citenamefont
  {Agrawal}(2003)}]{kivshar2003optical}%
  \BibitemOpen
  \bibfield  {author} {\bibinfo {author} {\bibfnamefont {Y.~S.}\ \bibnamefont
  {Kivshar}}\ and\ \bibinfo {author} {\bibfnamefont {G.~P.}\ \bibnamefont
  {Agrawal}},\ }\href@noop {} {\emph {\bibinfo {title} {Optical solitons: from
  fibers to photonic crystals}}}\ (\bibinfo  {publisher} {Academic press},\
  \bibinfo {address} {San Diego, California},\ \bibinfo {year}
  {2003})\BibitemShut {NoStop}%
\bibitem [{\citenamefont {Bock}\ \emph {et~al.}(2025)\citenamefont {Bock},
  \citenamefont {Ueberschaer}, \citenamefont {Steinmeyer},\ and\ \citenamefont
  {Griebner}}]{Bock2025}%
  \BibitemOpen
  \bibfield  {author} {\bibinfo {author} {\bibfnamefont {M.}~\bibnamefont
  {Bock}}, \bibinfo {author} {\bibfnamefont {D.}~\bibnamefont {Ueberschaer}},
  \bibinfo {author} {\bibfnamefont {G.}~\bibnamefont {Steinmeyer}},\ and\
  \bibinfo {author} {\bibfnamefont {U.}~\bibnamefont {Griebner}},\ }\bibfield
  {title} {\bibinfo {title} {Generation of two-cycle light bullets with 2.0 mj
  energy via self-compression of 5-µm pulses in zinc sulfide},\ }\href@noop {}
  {\bibfield  {journal} {\bibinfo  {journal} {Laser \& Photonics Reviews}\
  }\textbf {\bibinfo {volume} {19}},\ \bibinfo {pages} {e01102} (\bibinfo
  {year} {2025})}\BibitemShut {NoStop}%
\bibitem [{\citenamefont {Sazonov}(2025)}]{Sazonov2025}%
  \BibitemOpen
  \bibfield  {author} {\bibinfo {author} {\bibfnamefont {S.~V.}\ \bibnamefont
  {Sazonov}},\ }\bibfield  {title} {\bibinfo {title} {Two-color bright--dark
  light bullets at the presence of quasi-resonant impurities},\ }\href
  {https://doi.org/10.1088/1612-202X/ae18b9} {\bibfield  {journal} {\bibinfo
  {journal} {Laser Physics Letters}\ }\textbf {\bibinfo {volume} {22}},\
  \bibinfo {pages} {115201} (\bibinfo {year} {2025})}\BibitemShut {NoStop}%
\bibitem [{\citenamefont {B\'{e}jot}(2025)}]{Bejot2025}%
  \BibitemOpen
  \bibfield  {author} {\bibinfo {author} {\bibfnamefont {P.}~\bibnamefont
  {B\'{e}jot}},\ }\bibfield  {title} {\bibinfo {title} {Space-time multimode
  helical light bullets in ring-core fibers},\ }\href
  {https://doi.org/10.1364/JOSAB.562974} {\bibfield  {journal} {\bibinfo
  {journal} {J. Opt. Soc. Am. B}\ }\textbf {\bibinfo {volume} {42}},\ \bibinfo
  {pages} {1430} (\bibinfo {year} {2025})}\BibitemShut {NoStop}%
\bibitem [{\citenamefont {Chen}\ \emph {et~al.}(2023)\citenamefont {Chen},
  \citenamefont {Wu},\ and\ \citenamefont {Zeng}}]{Chen2023}%
  \BibitemOpen
  \bibfield  {author} {\bibinfo {author} {\bibfnamefont {Z.}~\bibnamefont
  {Chen}}, \bibinfo {author} {\bibfnamefont {Z.}~\bibnamefont {Wu}},\ and\
  \bibinfo {author} {\bibfnamefont {J.}~\bibnamefont {Zeng}},\ }\bibfield
  {title} {\bibinfo {title} {Light gap bullets in defocusing media with optical
  lattices},\ }\href
  {https://doi.org/https://doi.org/10.1016/j.chaos.2023.113785} {\bibfield
  {journal} {\bibinfo  {journal} {Chaos, Solitons \& Fractals}\ }\textbf
  {\bibinfo {volume} {174}},\ \bibinfo {pages} {113785} (\bibinfo {year}
  {2023})}\BibitemShut {NoStop}%
\bibitem [{\citenamefont {Aceves}\ \emph {et~al.}(2015)\citenamefont {Aceves},
  \citenamefont {Shtyrina}, \citenamefont {Rubenchik}, \citenamefont
  {Fedoruk},\ and\ \citenamefont {Turitsyn}}]{Aceves2015}%
  \BibitemOpen
  \bibfield  {author} {\bibinfo {author} {\bibfnamefont {A.~B.}\ \bibnamefont
  {Aceves}}, \bibinfo {author} {\bibfnamefont {O.~V.}\ \bibnamefont
  {Shtyrina}}, \bibinfo {author} {\bibfnamefont {A.~M.}\ \bibnamefont
  {Rubenchik}}, \bibinfo {author} {\bibfnamefont {M.~P.}\ \bibnamefont
  {Fedoruk}},\ and\ \bibinfo {author} {\bibfnamefont {S.~K.}\ \bibnamefont
  {Turitsyn}},\ }\bibfield  {title} {\bibinfo {title} {Spatiotemporal optical
  bullets in two-dimensional fiber arrays and their stability},\ }\href
  {https://doi.org/10.1103/PhysRevA.91.033810} {\bibfield  {journal} {\bibinfo
  {journal} {Phys. Rev. A}\ }\textbf {\bibinfo {volume} {91}},\ \bibinfo
  {pages} {033810} (\bibinfo {year} {2015})}\BibitemShut {NoStop}%
\bibitem [{\citenamefont {Wang}\ \emph {et~al.}(2025)\citenamefont {Wang},
  \citenamefont {Guo},\ and\ \citenamefont {Fan}}]{Wang2025}%
  \BibitemOpen
  \bibfield  {author} {\bibinfo {author} {\bibfnamefont {H.}~\bibnamefont
  {Wang}}, \bibinfo {author} {\bibfnamefont {C.}~\bibnamefont {Guo}},\ and\
  \bibinfo {author} {\bibfnamefont {S.}~\bibnamefont {Fan}},\ }\bibfield
  {title} {\bibinfo {title} {Spatiotemporal steering of nondiffracting wave
  packets},\ }\href {https://doi.org/10.1103/PhysRevLett.134.073803} {\bibfield
   {journal} {\bibinfo  {journal} {Phys. Rev. Lett.}\ }\textbf {\bibinfo
  {volume} {134}},\ \bibinfo {pages} {073803} (\bibinfo {year}
  {2025})}\BibitemShut {NoStop}%
\bibitem [{\citenamefont {Han}\ \emph {et~al.}(2026)\citenamefont {Han},
  \citenamefont {Li}, \citenamefont {Ge}, \citenamefont {Man}, \citenamefont
  {Fu}, \citenamefont {Wang}, \citenamefont {Chen}, \citenamefont {Wan},
  \citenamefont {Li},\ and\ \citenamefont {Zhang}}]{Han2026}%
  \BibitemOpen
  \bibfield  {author} {\bibinfo {author} {\bibfnamefont {X.}~\bibnamefont
  {Han}}, \bibinfo {author} {\bibfnamefont {S.}~\bibnamefont {Li}}, \bibinfo
  {author} {\bibfnamefont {X.}~\bibnamefont {Ge}}, \bibinfo {author}
  {\bibfnamefont {Z.}~\bibnamefont {Man}}, \bibinfo {author} {\bibfnamefont
  {S.}~\bibnamefont {Fu}}, \bibinfo {author} {\bibfnamefont {B.}~\bibnamefont
  {Wang}}, \bibinfo {author} {\bibfnamefont {C.}~\bibnamefont {Chen}}, \bibinfo
  {author} {\bibfnamefont {T.}~\bibnamefont {Wan}}, \bibinfo {author}
  {\bibfnamefont {H.}~\bibnamefont {Li}},\ and\ \bibinfo {author}
  {\bibfnamefont {L.}~\bibnamefont {Zhang}},\ }\bibfield  {title} {\bibinfo
  {title} {Nonlinear propagation characteristics of space-time circular
  airyprime wave packets in a kerr-type noninstantaneous media},\ }\href
  {https://doi.org/https://doi.org/10.1016/j.optlastec.2026.114742} {\bibfield
  {journal} {\bibinfo  {journal} {Optics \& Laser Technology}\ }\textbf
  {\bibinfo {volume} {197}},\ \bibinfo {pages} {114742} (\bibinfo {year}
  {2026})}\BibitemShut {NoStop}%
\bibitem [{\citenamefont {Fibich}(2015)}]{fibichnonlinear}%
  \BibitemOpen
  \bibfield  {author} {\bibinfo {author} {\bibfnamefont {G.}~\bibnamefont
  {Fibich}},\ }\href@noop {} {\emph {\bibinfo {title} {The Nonlinear
  Schr{\"o}dinger Equation: Singular Solutions and Optical Collapse}}},\
  Applied Mathematical Sciences\ (\bibinfo  {publisher} {Springer International
  Publishing},\ \bibinfo {address} {Switzerland},\ \bibinfo {year}
  {2015})\BibitemShut {NoStop}%
\bibitem [{\citenamefont {Bespalov}\ and\ \citenamefont
  {Talanov}(1966)}]{bespalov1966}%
  \BibitemOpen
  \bibfield  {author} {\bibinfo {author} {\bibfnamefont {V.~I.}\ \bibnamefont
  {Bespalov}}\ and\ \bibinfo {author} {\bibfnamefont {V.~I.}\ \bibnamefont
  {Talanov}},\ }\bibfield  {title} {\bibinfo {title} {Filamentary structure of
  light beams in nonlinear liquids},\ }\href@noop {} {\bibfield  {journal}
  {\bibinfo  {journal} {Soviet Journal of Experimental and Theoretical Physics
  Letters}\ }\textbf {\bibinfo {volume} {3}},\ \bibinfo {pages} {307} (\bibinfo
  {year} {1966})}\BibitemShut {NoStop}%
\bibitem [{\citenamefont {Sulem}\ and\ \citenamefont
  {Sulem}(2007)}]{sulem2007nonlinear}%
  \BibitemOpen
  \bibfield  {author} {\bibinfo {author} {\bibfnamefont {C.}~\bibnamefont
  {Sulem}}\ and\ \bibinfo {author} {\bibfnamefont {P.-L.}\ \bibnamefont
  {Sulem}},\ }\href@noop {} {\emph {\bibinfo {title} {The nonlinear
  Schr{\"o}dinger equation: self-focusing and wave collapse}}},\ Vol.\ \bibinfo
  {volume} {139}\ (\bibinfo  {publisher} {Springer Science \& Business Media},\
  \bibinfo {address} {New York},\ \bibinfo {year} {2007})\BibitemShut {NoStop}%
\bibitem [{\citenamefont {Edmundson}(1997)}]{edmundson1997unstable}%
  \BibitemOpen
  \bibfield  {author} {\bibinfo {author} {\bibfnamefont {D.~E.}\ \bibnamefont
  {Edmundson}},\ }\bibfield  {title} {\bibinfo {title} {Unstable higher modes
  of a three-dimensional nonlinear schr{\"o}dinger equation},\ }\href@noop {}
  {\bibfield  {journal} {\bibinfo  {journal} {Physical Review E}\ }\textbf
  {\bibinfo {volume} {55}},\ \bibinfo {pages} {7636} (\bibinfo {year}
  {1997})}\BibitemShut {NoStop}%
\bibitem [{\citenamefont {Malomed}(2022)}]{Malomed2022}%
  \BibitemOpen
  \bibfield  {author} {\bibinfo {author} {\bibfnamefont {B.~A.}\ \bibnamefont
  {Malomed}},\ }\href@noop {} {\emph {\bibinfo {title} {Multidimensional
  solitons}}}\ (\bibinfo  {publisher} {AIP Publishing},\ \bibinfo {address}
  {Melville, New York},\ \bibinfo {year} {2022})\BibitemShut {NoStop}%
\bibitem [{\citenamefont {Akhmediev}\ \emph {et~al.}(1992)\citenamefont
  {Akhmediev}, \citenamefont {Korneev},\ and\ \citenamefont
  {Nabiev}}]{akhmediev1992modulation}%
  \BibitemOpen
  \bibfield  {author} {\bibinfo {author} {\bibfnamefont {N.~N.}\ \bibnamefont
  {Akhmediev}}, \bibinfo {author} {\bibfnamefont {V.}~\bibnamefont {Korneev}},\
  and\ \bibinfo {author} {\bibfnamefont {R.~F.}\ \bibnamefont {Nabiev}},\
  }\bibfield  {title} {\bibinfo {title} {Modulation instability of the ground
  state of the nonlinear wave equation: optical machine gun},\ }\href@noop {}
  {\bibfield  {journal} {\bibinfo  {journal} {Optics letters}\ }\textbf
  {\bibinfo {volume} {17}},\ \bibinfo {pages} {393} (\bibinfo {year}
  {1992})}\BibitemShut {NoStop}%
\bibitem [{\citenamefont {Edmundson}\ and\ \citenamefont
  {Enns}(1992)}]{edmundson1992}%
  \BibitemOpen
  \bibfield  {author} {\bibinfo {author} {\bibfnamefont {D.}~\bibnamefont
  {Edmundson}}\ and\ \bibinfo {author} {\bibfnamefont {R.}~\bibnamefont
  {Enns}},\ }\bibfield  {title} {\bibinfo {title} {Robust bistable light
  bullets},\ }\href@noop {} {\bibfield  {journal} {\bibinfo  {journal} {Optics
  Letters}\ }\textbf {\bibinfo {volume} {17}},\ \bibinfo {pages} {586}
  (\bibinfo {year} {1992})}\BibitemShut {NoStop}%
\bibitem [{\citenamefont {Michinel}\ \emph {et~al.}(2002)\citenamefont
  {Michinel}, \citenamefont {Campo-T{\'a}boas}, \citenamefont
  {Garc{\'\i}a-Fern{\'a}ndez}, \citenamefont {Salgueiro},\ and\ \citenamefont
  {Quiroga-Teixeiro}}]{michinel2002liquid}%
  \BibitemOpen
  \bibfield  {author} {\bibinfo {author} {\bibfnamefont {H.}~\bibnamefont
  {Michinel}}, \bibinfo {author} {\bibfnamefont {J.}~\bibnamefont
  {Campo-T{\'a}boas}}, \bibinfo {author} {\bibfnamefont {R.}~\bibnamefont
  {Garc{\'\i}a-Fern{\'a}ndez}}, \bibinfo {author} {\bibfnamefont
  {J.}~\bibnamefont {Salgueiro}},\ and\ \bibinfo {author} {\bibfnamefont
  {M.}~\bibnamefont {Quiroga-Teixeiro}},\ }\bibfield  {title} {\bibinfo {title}
  {Liquid light condensates},\ }\href@noop {} {\bibfield  {journal} {\bibinfo
  {journal} {Physical Review E}\ }\textbf {\bibinfo {volume} {65}},\ \bibinfo
  {pages} {066604} (\bibinfo {year} {2002})}\BibitemShut {NoStop}%
\bibitem [{\citenamefont {Michinel}\ \emph {et~al.}(2006)\citenamefont
  {Michinel}, \citenamefont {Paz-Alonso},\ and\ \citenamefont
  {P{\'e}rez-Garc{\'\i}a}}]{michinel2006turning}%
  \BibitemOpen
  \bibfield  {author} {\bibinfo {author} {\bibfnamefont {H.}~\bibnamefont
  {Michinel}}, \bibinfo {author} {\bibfnamefont {M.~J.}\ \bibnamefont
  {Paz-Alonso}},\ and\ \bibinfo {author} {\bibfnamefont {V.~M.}\ \bibnamefont
  {P{\'e}rez-Garc{\'\i}a}},\ }\bibfield  {title} {\bibinfo {title} {Turning
  light into a liquid via atomic coherence},\ }\href@noop {} {\bibfield
  {journal} {\bibinfo  {journal} {Phys. Rev. Lett.}\ }\textbf {\bibinfo
  {volume} {96}},\ \bibinfo {pages} {023903} (\bibinfo {year}
  {2006})}\BibitemShut {NoStop}%
\bibitem [{\citenamefont {Krolikowski}\ \emph {et~al.}(2001)\citenamefont
  {Krolikowski}, \citenamefont {Bang}, \citenamefont {Rasmussen},\ and\
  \citenamefont {Wyller}}]{krolikowski2001modulational}%
  \BibitemOpen
  \bibfield  {author} {\bibinfo {author} {\bibfnamefont {W.}~\bibnamefont
  {Krolikowski}}, \bibinfo {author} {\bibfnamefont {O.}~\bibnamefont {Bang}},
  \bibinfo {author} {\bibfnamefont {J.~J.}\ \bibnamefont {Rasmussen}},\ and\
  \bibinfo {author} {\bibfnamefont {J.}~\bibnamefont {Wyller}},\ }\bibfield
  {title} {\bibinfo {title} {Modulational instability in nonlocal nonlinear
  kerr media},\ }\href@noop {} {\bibfield  {journal} {\bibinfo  {journal}
  {Physical Review E}\ }\textbf {\bibinfo {volume} {64}},\ \bibinfo {pages}
  {016612} (\bibinfo {year} {2001})}\BibitemShut {NoStop}%
\bibitem [{\citenamefont {Bang}\ \emph {et~al.}(2002)\citenamefont {Bang},
  \citenamefont {Krolikowski}, \citenamefont {Wyller},\ and\ \citenamefont
  {Rasmussen}}]{bang2002collapse}%
  \BibitemOpen
  \bibfield  {author} {\bibinfo {author} {\bibfnamefont {O.}~\bibnamefont
  {Bang}}, \bibinfo {author} {\bibfnamefont {W.}~\bibnamefont {Krolikowski}},
  \bibinfo {author} {\bibfnamefont {J.}~\bibnamefont {Wyller}},\ and\ \bibinfo
  {author} {\bibfnamefont {J.~J.}\ \bibnamefont {Rasmussen}},\ }\bibfield
  {title} {\bibinfo {title} {Collapse arrest and soliton stabilization in
  nonlocal nonlinear media},\ }\href@noop {} {\bibfield  {journal} {\bibinfo
  {journal} {Physical Review E}\ }\textbf {\bibinfo {volume} {66}},\ \bibinfo
  {pages} {046619} (\bibinfo {year} {2002})}\BibitemShut {NoStop}%
\bibitem [{\citenamefont {Mihalache}\ \emph {et~al.}(2006)\citenamefont
  {Mihalache}, \citenamefont {Mazilu}, \citenamefont {Lederer}, \citenamefont
  {Malomed}, \citenamefont {Kartashov}, \citenamefont {Crasovan},\ and\
  \citenamefont {Torner}}]{mihalache2006three}%
  \BibitemOpen
  \bibfield  {author} {\bibinfo {author} {\bibfnamefont {D.}~\bibnamefont
  {Mihalache}}, \bibinfo {author} {\bibfnamefont {D.}~\bibnamefont {Mazilu}},
  \bibinfo {author} {\bibfnamefont {F.}~\bibnamefont {Lederer}}, \bibinfo
  {author} {\bibfnamefont {B.}~\bibnamefont {Malomed}}, \bibinfo {author}
  {\bibfnamefont {Y.~V.}\ \bibnamefont {Kartashov}}, \bibinfo {author}
  {\bibfnamefont {L.-C.}\ \bibnamefont {Crasovan}},\ and\ \bibinfo {author}
  {\bibfnamefont {L.}~\bibnamefont {Torner}},\ }\bibfield  {title} {\bibinfo
  {title} {Three-dimensional spatiotemporal optical solitons in nonlocal
  nonlinear media},\ }\href@noop {} {\bibfield  {journal} {\bibinfo  {journal}
  {Physical Review E---Statistical, Nonlinear, and Soft Matter Physics}\
  }\textbf {\bibinfo {volume} {73}},\ \bibinfo {pages} {025601} (\bibinfo
  {year} {2006})}\BibitemShut {NoStop}%
\bibitem [{\citenamefont {Paredes}\ \emph {et~al.}(2020)\citenamefont
  {Paredes}, \citenamefont {Olivieri},\ and\ \citenamefont
  {Michinel}}]{paredes2020optics}%
  \BibitemOpen
  \bibfield  {author} {\bibinfo {author} {\bibfnamefont {A.}~\bibnamefont
  {Paredes}}, \bibinfo {author} {\bibfnamefont {D.~N.}\ \bibnamefont
  {Olivieri}},\ and\ \bibinfo {author} {\bibfnamefont {H.}~\bibnamefont
  {Michinel}},\ }\bibfield  {title} {\bibinfo {title} {From optics to dark
  matter: A review on nonlinear schr{\"o}dinger--poisson systems},\ }\href@noop
  {} {\bibfield  {journal} {\bibinfo  {journal} {Physica D: Nonlinear
  Phenomena}\ }\textbf {\bibinfo {volume} {403}},\ \bibinfo {pages} {132301}
  (\bibinfo {year} {2020})}\BibitemShut {NoStop}%
\bibitem [{\citenamefont {Baizakov}\ \emph {et~al.}(2003)\citenamefont
  {Baizakov}, \citenamefont {Malomed},\ and\ \citenamefont
  {Salerno}}]{baizakov2003multidimensional}%
  \BibitemOpen
  \bibfield  {author} {\bibinfo {author} {\bibfnamefont {B.~B.}\ \bibnamefont
  {Baizakov}}, \bibinfo {author} {\bibfnamefont {B.~A.}\ \bibnamefont
  {Malomed}},\ and\ \bibinfo {author} {\bibfnamefont {M.}~\bibnamefont
  {Salerno}},\ }\bibfield  {title} {\bibinfo {title} {Multidimensional solitons
  in periodic potentials},\ }\href@noop {} {\bibfield  {journal} {\bibinfo
  {journal} {Europhysics letters}\ }\textbf {\bibinfo {volume} {63}},\ \bibinfo
  {pages} {642} (\bibinfo {year} {2003})}\BibitemShut {NoStop}%
\bibitem [{\citenamefont {Mihalache}\ \emph {et~al.}(2005)\citenamefont
  {Mihalache}, \citenamefont {Mazilu}, \citenamefont {Lederer}, \citenamefont
  {Malomed}, \citenamefont {Kartashov}, \citenamefont {Crasovan},\ and\
  \citenamefont {Torner}}]{mihalache2005stable}%
  \BibitemOpen
  \bibfield  {author} {\bibinfo {author} {\bibfnamefont {D.}~\bibnamefont
  {Mihalache}}, \bibinfo {author} {\bibfnamefont {D.}~\bibnamefont {Mazilu}},
  \bibinfo {author} {\bibfnamefont {F.}~\bibnamefont {Lederer}}, \bibinfo
  {author} {\bibfnamefont {B.}~\bibnamefont {Malomed}}, \bibinfo {author}
  {\bibfnamefont {Y.~V.}\ \bibnamefont {Kartashov}}, \bibinfo {author}
  {\bibfnamefont {L.-C.}\ \bibnamefont {Crasovan}},\ and\ \bibinfo {author}
  {\bibfnamefont {L.}~\bibnamefont {Torner}},\ }\bibfield  {title} {\bibinfo
  {title} {Stable spatiotemporal solitons in bessel optical lattices},\
  }\href@noop {} {\bibfield  {journal} {\bibinfo  {journal} {Physical review
  letters}\ }\textbf {\bibinfo {volume} {95}},\ \bibinfo {pages} {023902}
  (\bibinfo {year} {2005})}\BibitemShut {NoStop}%
\bibitem [{\citenamefont {Diebel}\ \emph {et~al.}(2016)\citenamefont {Diebel},
  \citenamefont {Boguslawski}, \citenamefont {Dadalyan}, \citenamefont
  {Drampyan},\ and\ \citenamefont {Denz}}]{diebel2016controlled}%
  \BibitemOpen
  \bibfield  {author} {\bibinfo {author} {\bibfnamefont {F.}~\bibnamefont
  {Diebel}}, \bibinfo {author} {\bibfnamefont {M.}~\bibnamefont {Boguslawski}},
  \bibinfo {author} {\bibfnamefont {T.}~\bibnamefont {Dadalyan}}, \bibinfo
  {author} {\bibfnamefont {R.}~\bibnamefont {Drampyan}},\ and\ \bibinfo
  {author} {\bibfnamefont {C.}~\bibnamefont {Denz}},\ }\bibfield  {title}
  {\bibinfo {title} {Controlled soliton formation in tailored bessel photonic
  lattices},\ }\href@noop {} {\bibfield  {journal} {\bibinfo  {journal} {Optics
  Express}\ }\textbf {\bibinfo {volume} {24}},\ \bibinfo {pages} {12933}
  (\bibinfo {year} {2016})}\BibitemShut {NoStop}%
\bibitem [{\citenamefont {Ak{\"o}zbek}\ and\ \citenamefont
  {John}(1998)}]{akozbek1998optical}%
  \BibitemOpen
  \bibfield  {author} {\bibinfo {author} {\bibfnamefont {N.}~\bibnamefont
  {Ak{\"o}zbek}}\ and\ \bibinfo {author} {\bibfnamefont {S.}~\bibnamefont
  {John}},\ }\bibfield  {title} {\bibinfo {title} {Optical solitary waves in
  two-and three-dimensional nonlinear photonic band-gap structures},\
  }\href@noop {} {\bibfield  {journal} {\bibinfo  {journal} {Physical Review
  E}\ }\textbf {\bibinfo {volume} {57}},\ \bibinfo {pages} {2287} (\bibinfo
  {year} {1998})}\BibitemShut {NoStop}%
\bibitem [{\citenamefont {Kartashov}(2022)}]{Kartashov2022}%
  \BibitemOpen
  \bibfield  {author} {\bibinfo {author} {\bibfnamefont {Y.~V.}\ \bibnamefont
  {Kartashov}},\ }\bibfield  {title} {\bibinfo {title} {Light bullets in
  moir\'e lattices},\ }\href {https://doi.org/10.1364/OL.471022} {\bibfield
  {journal} {\bibinfo  {journal} {Opt. Lett.}\ }\textbf {\bibinfo {volume}
  {47}},\ \bibinfo {pages} {4528} (\bibinfo {year} {2022})}\BibitemShut
  {NoStop}%
\bibitem [{\citenamefont {Mili\'an}\ \emph {et~al.}(2019)\citenamefont
  {Mili\'an}, \citenamefont {Kartashov},\ and\ \citenamefont
  {Torner}}]{Milian2019}%
  \BibitemOpen
  \bibfield  {author} {\bibinfo {author} {\bibfnamefont {C.}~\bibnamefont
  {Mili\'an}}, \bibinfo {author} {\bibfnamefont {Y.~V.}\ \bibnamefont
  {Kartashov}},\ and\ \bibinfo {author} {\bibfnamefont {L.}~\bibnamefont
  {Torner}},\ }\bibfield  {title} {\bibinfo {title} {Robust ultrashort light
  bullets in strongly twisted waveguide arrays},\ }\href
  {https://doi.org/10.1103/PhysRevLett.123.133902} {\bibfield  {journal}
  {\bibinfo  {journal} {Phys. Rev. Lett.}\ }\textbf {\bibinfo {volume} {123}},\
  \bibinfo {pages} {133902} (\bibinfo {year} {2019})}\BibitemShut {NoStop}%
\bibitem [{\citenamefont {Nye}\ and\ \citenamefont
  {Berry}(1974)}]{nye1974dislocations}%
  \BibitemOpen
  \bibfield  {author} {\bibinfo {author} {\bibfnamefont {J.~F.}\ \bibnamefont
  {Nye}}\ and\ \bibinfo {author} {\bibfnamefont {M.~V.}\ \bibnamefont
  {Berry}},\ }\bibfield  {title} {\bibinfo {title} {Dislocations in wave
  trains},\ }\href@noop {} {\bibfield  {journal} {\bibinfo  {journal} {Proc.
  Math. Phys. Eng. Sci.}\ }\textbf {\bibinfo {volume} {336}},\ \bibinfo {pages}
  {165} (\bibinfo {year} {1974})}\BibitemShut {NoStop}%
\bibitem [{\citenamefont {Coullet}\ \emph {et~al.}(1989)\citenamefont
  {Coullet}, \citenamefont {Gil},\ and\ \citenamefont {Rocca}}]{coullet1989}%
  \BibitemOpen
  \bibfield  {author} {\bibinfo {author} {\bibfnamefont {P.}~\bibnamefont
  {Coullet}}, \bibinfo {author} {\bibfnamefont {L.}~\bibnamefont {Gil}},\ and\
  \bibinfo {author} {\bibfnamefont {F.}~\bibnamefont {Rocca}},\ }\bibfield
  {title} {\bibinfo {title} {Optical vortices},\ }\href@noop {} {\bibfield
  {journal} {\bibinfo  {journal} {Optics Communications}\ }\textbf {\bibinfo
  {volume} {73}},\ \bibinfo {pages} {403} (\bibinfo {year} {1989})}\BibitemShut
  {NoStop}%
\bibitem [{\citenamefont {Desyatnikov}\ \emph
  {et~al.}(2005{\natexlab{a}})\citenamefont {Desyatnikov}, \citenamefont
  {Sukhorukov},\ and\ \citenamefont {Kivshar}}]{desyatnikov2005azimuthons}%
  \BibitemOpen
  \bibfield  {author} {\bibinfo {author} {\bibfnamefont {A.~S.}\ \bibnamefont
  {Desyatnikov}}, \bibinfo {author} {\bibfnamefont {A.~A.}\ \bibnamefont
  {Sukhorukov}},\ and\ \bibinfo {author} {\bibfnamefont {Y.~S.}\ \bibnamefont
  {Kivshar}},\ }\bibfield  {title} {\bibinfo {title} {Azimuthons: spatially
  modulated vortex solitons},\ }\href@noop {} {\bibfield  {journal} {\bibinfo
  {journal} {Physical review letters}\ }\textbf {\bibinfo {volume} {95}},\
  \bibinfo {pages} {203904} (\bibinfo {year} {2005}{\natexlab{a}})}\BibitemShut
  {NoStop}%
\bibitem [{\citenamefont {Desyatnikov}\ \emph
  {et~al.}(2005{\natexlab{b}})\citenamefont {Desyatnikov}, \citenamefont
  {Kivshar},\ and\ \citenamefont {Torner}}]{desyatnikov2005optical}%
  \BibitemOpen
  \bibfield  {author} {\bibinfo {author} {\bibfnamefont {A.~S.}\ \bibnamefont
  {Desyatnikov}}, \bibinfo {author} {\bibfnamefont {Y.~S.}\ \bibnamefont
  {Kivshar}},\ and\ \bibinfo {author} {\bibfnamefont {L.}~\bibnamefont
  {Torner}},\ }\bibfield  {title} {\bibinfo {title} {Optical vortices and
  vortex solitons},\ }in\ \href@noop {} {\emph {\bibinfo {booktitle} {Progress
  in optics}}},\ Vol.~\bibinfo {volume} {47}\ (\bibinfo  {publisher}
  {Elsevier},\ \bibinfo {address} {{}},\ \bibinfo {year} {2005})\ pp.\ \bibinfo
  {pages} {291--391}\BibitemShut {NoStop}%
\bibitem [{\citenamefont {Quiroga-Teixeiro}\ and\ \citenamefont
  {Michinel}(1997)}]{quiroga1997}%
  \BibitemOpen
  \bibfield  {author} {\bibinfo {author} {\bibfnamefont {M.}~\bibnamefont
  {Quiroga-Teixeiro}}\ and\ \bibinfo {author} {\bibfnamefont {H.}~\bibnamefont
  {Michinel}},\ }\bibfield  {title} {\bibinfo {title} {Stable azimuthal
  stationary state in quintic nonlinear optical media},\ }\href@noop {}
  {\bibfield  {journal} {\bibinfo  {journal} {J. Opt. Soc. Am. B}\ }\textbf
  {\bibinfo {volume} {14}},\ \bibinfo {pages} {2004} (\bibinfo {year}
  {1997})}\BibitemShut {NoStop}%
\bibitem [{\citenamefont {Michinel}\ \emph {et~al.}(2004)\citenamefont
  {Michinel}, \citenamefont {Salgueiro},\ and\ \citenamefont
  {Paz-Alonso}}]{michinel2004square}%
  \BibitemOpen
  \bibfield  {author} {\bibinfo {author} {\bibfnamefont {H.}~\bibnamefont
  {Michinel}}, \bibinfo {author} {\bibfnamefont {J.~R.}\ \bibnamefont
  {Salgueiro}},\ and\ \bibinfo {author} {\bibfnamefont {M.~J.}\ \bibnamefont
  {Paz-Alonso}},\ }\bibfield  {title} {\bibinfo {title} {Square vortex solitons
  with a large angular momentum},\ }\href@noop {} {\bibfield  {journal}
  {\bibinfo  {journal} {Phys. Rev. E}\ }\textbf {\bibinfo {volume} {70}},\
  \bibinfo {pages} {066605} (\bibinfo {year} {2004})}\BibitemShut {NoStop}%
\bibitem [{\citenamefont {Salamin}(2018)}]{salamin2018fields}%
  \BibitemOpen
  \bibfield  {author} {\bibinfo {author} {\bibfnamefont {Y.~I.}\ \bibnamefont
  {Salamin}},\ }\bibfield  {title} {\bibinfo {title} {Fields of a bessel-bessel
  light bullet of arbitrary order in an under-dense plasma},\ }\href@noop {}
  {\bibfield  {journal} {\bibinfo  {journal} {Scientific Reports}\ }\textbf
  {\bibinfo {volume} {8}},\ \bibinfo {pages} {11362} (\bibinfo {year}
  {2018})}\BibitemShut {NoStop}%
\bibitem [{\citenamefont {Hang}\ and\ \citenamefont
  {Huang}(2013)}]{hang2013ultraslow}%
  \BibitemOpen
  \bibfield  {author} {\bibinfo {author} {\bibfnamefont {C.}~\bibnamefont
  {Hang}}\ and\ \bibinfo {author} {\bibfnamefont {G.}~\bibnamefont {Huang}},\
  }\bibfield  {title} {\bibinfo {title} {Ultraslow helical optical bullets and
  their acceleration in magneto-optically controlled coherent atomic media},\
  }\href@noop {} {\bibfield  {journal} {\bibinfo  {journal} {Physical Review
  A---Atomic, Molecular, and Optical Physics}\ }\textbf {\bibinfo {volume}
  {87}},\ \bibinfo {pages} {053809} (\bibinfo {year} {2013})}\BibitemShut
  {NoStop}%
\bibitem [{\citenamefont {Porras}(2019)}]{porras2019upper}%
  \BibitemOpen
  \bibfield  {author} {\bibinfo {author} {\bibfnamefont {M.~A.}\ \bibnamefont
  {Porras}},\ }\bibfield  {title} {\bibinfo {title} {Upper bound to the orbital
  angular momentum carried by an ultrashort pulse},\ }\href@noop {} {\bibfield
  {journal} {\bibinfo  {journal} {Physical Review Letters}\ }\textbf {\bibinfo
  {volume} {122}},\ \bibinfo {pages} {123904} (\bibinfo {year}
  {2019})}\BibitemShut {NoStop}%
\bibitem [{\citenamefont {Chen}\ \emph {et~al.}(1997)\citenamefont {Chen},
  \citenamefont {Segev}, \citenamefont {Wilson}, \citenamefont {Muller},\ and\
  \citenamefont {Maker}}]{chen1997self}%
  \BibitemOpen
  \bibfield  {author} {\bibinfo {author} {\bibfnamefont {Z.}~\bibnamefont
  {Chen}}, \bibinfo {author} {\bibfnamefont {M.}~\bibnamefont {Segev}},
  \bibinfo {author} {\bibfnamefont {D.~W.}\ \bibnamefont {Wilson}}, \bibinfo
  {author} {\bibfnamefont {R.~E.}\ \bibnamefont {Muller}},\ and\ \bibinfo
  {author} {\bibfnamefont {P.~D.}\ \bibnamefont {Maker}},\ }\bibfield  {title}
  {\bibinfo {title} {Self-trapping of an optical vortex by use of the bulk
  photovoltaic effect},\ }\href@noop {} {\bibfield  {journal} {\bibinfo
  {journal} {Physical Review Letters}\ }\textbf {\bibinfo {volume} {78}},\
  \bibinfo {pages} {2948} (\bibinfo {year} {1997})}\BibitemShut {NoStop}%
\bibitem [{\citenamefont {Zhang}\ \emph {et~al.}(2022)\citenamefont {Zhang},
  \citenamefont {Zhou},\ and\ \citenamefont {Dai}}]{zhang2022stabilization}%
  \BibitemOpen
  \bibfield  {author} {\bibinfo {author} {\bibfnamefont {H.}~\bibnamefont
  {Zhang}}, \bibinfo {author} {\bibfnamefont {T.}~\bibnamefont {Zhou}},\ and\
  \bibinfo {author} {\bibfnamefont {C.}~\bibnamefont {Dai}},\ }\bibfield
  {title} {\bibinfo {title} {Stabilization of higher-order vortex solitons by
  means of nonlocal nonlinearity},\ }\href@noop {} {\bibfield  {journal}
  {\bibinfo  {journal} {Physical Review A}\ }\textbf {\bibinfo {volume}
  {105}},\ \bibinfo {pages} {013520} (\bibinfo {year} {2022})}\BibitemShut
  {NoStop}%
\bibitem [{\citenamefont {Eilenberger}\ \emph {et~al.}(2013)\citenamefont
  {Eilenberger}, \citenamefont {Prater}, \citenamefont {Minardi}, \citenamefont
  {Geiss}, \citenamefont {R\"opke}, \citenamefont {Kobelke}, \citenamefont
  {Schuster}, \citenamefont {Bartelt}, \citenamefont {Nolte}, \citenamefont
  {T\"unnermann},\ and\ \citenamefont {Pertsch}}]{Eilenberger2013}%
  \BibitemOpen
  \bibfield  {author} {\bibinfo {author} {\bibfnamefont {F.}~\bibnamefont
  {Eilenberger}}, \bibinfo {author} {\bibfnamefont {K.}~\bibnamefont {Prater}},
  \bibinfo {author} {\bibfnamefont {S.}~\bibnamefont {Minardi}}, \bibinfo
  {author} {\bibfnamefont {R.}~\bibnamefont {Geiss}}, \bibinfo {author}
  {\bibfnamefont {U.}~\bibnamefont {R\"opke}}, \bibinfo {author} {\bibfnamefont
  {J.}~\bibnamefont {Kobelke}}, \bibinfo {author} {\bibfnamefont
  {K.}~\bibnamefont {Schuster}}, \bibinfo {author} {\bibfnamefont
  {H.}~\bibnamefont {Bartelt}}, \bibinfo {author} {\bibfnamefont
  {S.}~\bibnamefont {Nolte}}, \bibinfo {author} {\bibfnamefont
  {A.}~\bibnamefont {T\"unnermann}},\ and\ \bibinfo {author} {\bibfnamefont
  {T.}~\bibnamefont {Pertsch}},\ }\bibfield  {title} {\bibinfo {title}
  {Observation of discrete, vortex light bullets},\ }\href
  {https://doi.org/10.1103/PhysRevX.3.041031} {\bibfield  {journal} {\bibinfo
  {journal} {Phys. Rev. X}\ }\textbf {\bibinfo {volume} {3}},\ \bibinfo {pages}
  {041031} (\bibinfo {year} {2013})}\BibitemShut {NoStop}%
\bibitem [{\citenamefont {Zhang}\ \emph {et~al.}(2026)\citenamefont {Zhang},
  \citenamefont {Ivanov}, \citenamefont {Li}, \citenamefont {Kartashov},\ and\
  \citenamefont {Zhang}}]{Zhang2026}%
  \BibitemOpen
  \bibfield  {author} {\bibinfo {author} {\bibfnamefont {Z.}~\bibnamefont
  {Zhang}}, \bibinfo {author} {\bibfnamefont {S.~K.}\ \bibnamefont {Ivanov}},
  \bibinfo {author} {\bibfnamefont {Y.}~\bibnamefont {Li}}, \bibinfo {author}
  {\bibfnamefont {Y.~V.}\ \bibnamefont {Kartashov}},\ and\ \bibinfo {author}
  {\bibfnamefont {Y.}~\bibnamefont {Zhang}},\ }\bibfield  {title} {\bibinfo
  {title} {Vortex light bullets guided by disclination defects},\ }\href
  {https://doi.org/https://doi.org/10.1002/nap2.70206} {\bibfield  {journal}
  {\bibinfo  {journal} {Nanophotonics}\ }\textbf {\bibinfo {volume} {15}},\
  \bibinfo {pages} {e70206} (\bibinfo {year} {2026})},\ \Eprint
  {https://arxiv.org/abs/https://onlinelibrary.wiley.com/doi/pdf/10.1002/nap2.70206}
  {https://onlinelibrary.wiley.com/doi/pdf/10.1002/nap2.70206} \BibitemShut
  {NoStop}%
\bibitem [{\citenamefont {S\'anchez}\ \emph {et~al.}(2026)\citenamefont
  {S\'anchez}, \citenamefont {Paredes}, \citenamefont {Michinel}, \citenamefont
  {Malomed},\ and\ \citenamefont {Salgueiro}}]{Sanchez2026}%
  \BibitemOpen
  \bibfield  {author} {\bibinfo {author} {\bibfnamefont {C.~F.}\ \bibnamefont
  {S\'anchez}}, \bibinfo {author} {\bibfnamefont {A.}~\bibnamefont {Paredes}},
  \bibinfo {author} {\bibfnamefont {H.}~\bibnamefont {Michinel}}, \bibinfo
  {author} {\bibfnamefont {B.~A.}\ \bibnamefont {Malomed}},\ and\ \bibinfo
  {author} {\bibfnamefont {J.~R.}\ \bibnamefont {Salgueiro}},\ }\bibfield
  {title} {\bibinfo {title} {Guided vortex bullets},\ }\href
  {https://doi.org/10.1103/6mj1-yc1m} {\bibfield  {journal} {\bibinfo
  {journal} {Phys. Rev. Lett.}\ }\textbf {\bibinfo {volume} {136}},\ \bibinfo
  {pages} {033802} (\bibinfo {year} {2026})}\BibitemShut {NoStop}%
\bibitem [{\citenamefont {Dunjko}\ \emph {et~al.}(2001)\citenamefont {Dunjko},
  \citenamefont {Lorent},\ and\ \citenamefont {Olshanii}}]{Dunjko2001}%
  \BibitemOpen
  \bibfield  {author} {\bibinfo {author} {\bibfnamefont {V.}~\bibnamefont
  {Dunjko}}, \bibinfo {author} {\bibfnamefont {V.}~\bibnamefont {Lorent}},\
  and\ \bibinfo {author} {\bibfnamefont {M.}~\bibnamefont {Olshanii}},\
  }\bibfield  {title} {\bibinfo {title} {Bosons in cigar-shaped traps:
  Thomas-fermi regime, tonks-girardeau regime, and in between},\ }\href
  {https://doi.org/10.1103/PhysRevLett.86.5413} {\bibfield  {journal} {\bibinfo
   {journal} {Phys. Rev. Lett.}\ }\textbf {\bibinfo {volume} {86}},\ \bibinfo
  {pages} {5413} (\bibinfo {year} {2001})}\BibitemShut {NoStop}%
\bibitem [{\citenamefont {Salgueiro}\ \emph {et~al.}(2007)\citenamefont
  {Salgueiro}, \citenamefont {Olivieri},\ and\ \citenamefont
  {Michinel}}]{salgueiro2007computation}%
  \BibitemOpen
  \bibfield  {author} {\bibinfo {author} {\bibfnamefont {J.~R.}\ \bibnamefont
  {Salgueiro}}, \bibinfo {author} {\bibfnamefont {D.}~\bibnamefont
  {Olivieri}},\ and\ \bibinfo {author} {\bibfnamefont {H.}~\bibnamefont
  {Michinel}},\ }\bibfield  {title} {\bibinfo {title} {Computation of linear
  and nonlinear stationary states of photonic structures using modern iterative
  solvers},\ }\href@noop {} {\bibfield  {journal} {\bibinfo  {journal} {Optical
  and quantum electronics}\ }\textbf {\bibinfo {volume} {39}},\ \bibinfo
  {pages} {239} (\bibinfo {year} {2007})}\BibitemShut {NoStop}%
\bibitem [{\citenamefont {Vakhitov}\ and\ \citenamefont
  {Kolokolov}(1973)}]{Vakhitov1973}%
  \BibitemOpen
  \bibfield  {author} {\bibinfo {author} {\bibfnamefont {N.~G.}\ \bibnamefont
  {Vakhitov}}\ and\ \bibinfo {author} {\bibfnamefont {A.~A.}\ \bibnamefont
  {Kolokolov}},\ }\bibfield  {title} {\bibinfo {title} {Stationary solutions of
  the wave equation in a medium with nonlinearity saturation},\ }\href
  {https://doi.org/10.1007/BF01031343} {\bibfield  {journal} {\bibinfo
  {journal} {Radiophysics and Quantum Electronics}\ }\textbf {\bibinfo {volume}
  {16}},\ \bibinfo {pages} {783} (\bibinfo {year} {1973})}\BibitemShut
  {NoStop}%
\bibitem [{\citenamefont {Press}\ \emph {et~al.}(1992)\citenamefont {Press},
  \citenamefont {Teukolsky}, \citenamefont {Vetterling},\ and\ \citenamefont
  {Flannery}}]{Press1992}%
  \BibitemOpen
  \bibfield  {author} {\bibinfo {author} {\bibfnamefont {W.~H.}\ \bibnamefont
  {Press}}, \bibinfo {author} {\bibfnamefont {S.~A.}\ \bibnamefont
  {Teukolsky}}, \bibinfo {author} {\bibfnamefont {W.~T.}\ \bibnamefont
  {Vetterling}},\ and\ \bibinfo {author} {\bibfnamefont {B.~P.}\ \bibnamefont
  {Flannery}},\ }\href@noop {} {\emph {\bibinfo {title} {Numerical recipes in
  {C}}}},\ \bibinfo {edition} {2nd}\ ed.\ (\bibinfo  {publisher} {Cambridge
  University Press},\ \bibinfo {address} {Cambridge},\ \bibinfo {year}
  {1992})\BibitemShut {NoStop}%
\bibitem [{\citenamefont {Hadley}(1992)}]{Hadley1992}%
  \BibitemOpen
  \bibfield  {author} {\bibinfo {author} {\bibfnamefont {G.~R.}\ \bibnamefont
  {Hadley}},\ }\bibfield  {title} {\bibinfo {title} {Transparent boundary
  condition for the beam propagation method},\ }\href@noop {} {\bibfield
  {journal} {\bibinfo  {journal} {IEEE Journal of quantum electronics}\
  }\textbf {\bibinfo {volume} {28}},\ \bibinfo {pages} {363} (\bibinfo {year}
  {1992})}\BibitemShut {NoStop}%
\bibitem [{\citenamefont {Chang}\ and\ \citenamefont {Chow}(1991)}]{Chang1991}%
  \BibitemOpen
  \bibfield  {author} {\bibinfo {author} {\bibfnamefont {M.~J.}\ \bibnamefont
  {Chang}}\ and\ \bibinfo {author} {\bibfnamefont {L.~C.}\ \bibnamefont
  {Chow}},\ }\bibfield  {title} {\bibinfo {title} {Improved
  alternating-direction implicit method for solving transient three-dimensional
  heat diffusion problems},\ }\href@noop {} {\bibfield  {journal} {\bibinfo
  {journal} {Numerical Heat Transfer}\ }\textbf {\bibinfo {volume} {19}},\
  \bibinfo {pages} {69} (\bibinfo {year} {1991})}\BibitemShut {NoStop}%
\bibitem [{\citenamefont {Salgueiro}(2016)}]{Salgueiro2016}%
  \BibitemOpen
  \bibfield  {author} {\bibinfo {author} {\bibfnamefont {J.~R.}\ \bibnamefont
  {Salgueiro}},\ }\bibfield  {title} {\bibinfo {title} {Vector--vortex solitons
  in nonlinear photonic crystal fibers},\ }\href@noop {} {\bibfield  {journal}
  {\bibinfo  {journal} {J. Opt.}\ }\textbf {\bibinfo {volume} {18}},\ \bibinfo
  {pages} {074004} (\bibinfo {year} {2016})}\BibitemShut {NoStop}%
\bibitem [{\citenamefont {Izdebskaya}\ \emph {et~al.}(2012)\citenamefont
  {Izdebskaya}, \citenamefont {Rebling}, \citenamefont {Desyatnikov},\ and\
  \citenamefont {Kivshar}}]{Izdebskaya2012}%
  \BibitemOpen
  \bibfield  {author} {\bibinfo {author} {\bibfnamefont {Y.~V.}\ \bibnamefont
  {Izdebskaya}}, \bibinfo {author} {\bibfnamefont {J.}~\bibnamefont {Rebling}},
  \bibinfo {author} {\bibfnamefont {A.~S.}\ \bibnamefont {Desyatnikov}},\ and\
  \bibinfo {author} {\bibfnamefont {Y.~S.}\ \bibnamefont {Kivshar}},\
  }\bibfield  {title} {\bibinfo {title} {Observation of vector solitons with
  hidden vorticity},\ }\href@noop {} {\bibfield  {journal} {\bibinfo  {journal}
  {Opt. Lett.}\ }\textbf {\bibinfo {volume} {37}},\ \bibinfo {pages} {767}
  (\bibinfo {year} {2012})}\BibitemShut {NoStop}%
\bibitem [{\citenamefont {Brtka}\ \emph {et~al.}(2010)\citenamefont {Brtka},
  \citenamefont {Gammal},\ and\ \citenamefont {Malomed}}]{Brtka2010}%
  \BibitemOpen
  \bibfield  {author} {\bibinfo {author} {\bibfnamefont {M.}~\bibnamefont
  {Brtka}}, \bibinfo {author} {\bibfnamefont {A.}~\bibnamefont {Gammal}},\ and\
  \bibinfo {author} {\bibfnamefont {B.~A.}\ \bibnamefont {Malomed}},\
  }\bibfield  {title} {\bibinfo {title} {Hidden vorticity in binary
  {B}ose-{E}instein condensates},\ }\href
  {http://link.aps.org/doi/10.1103/PhysRevA.82.053610} {\bibfield  {journal}
  {\bibinfo  {journal} {Phys. Rev. A}\ }\textbf {\bibinfo {volume} {82}},\
  \bibinfo {pages} {053610} (\bibinfo {year} {2010})}\BibitemShut {NoStop}%
\end{thebibliography}%

\end{document}